\documentclass{article}

\usepackage[preprint,nonatbib]{neurips_2021}

\usepackage[utf8]{inputenc} 
\usepackage[T1]{fontenc}    
\usepackage{hyperref}       
\usepackage{url}            
\usepackage{booktabs}       
\usepackage{amsfonts}       
\usepackage{nicefrac}       
\usepackage{microtype}      
\usepackage{xcolor}         

\usepackage[toc,page,header]{appendix}
\usepackage{minitoc}

\doparttoc 
\faketableofcontents 
\definecolor{mydarkblue}{rgb}{0,0.08,0.45}
\usepackage{wrapfig}
\usepackage{subcaption}
\newlength{\twosubht}
\newsavebox{\twosubbox}
\usepackage{amsmath,amsthm,amscd,amsbsy,amssymb,latexsym,url,bm}
\newcommand{\bs}{\boldsymbol}

\usepackage{cuted}
\usepackage{flushend}

\usepackage{amsfonts}       
\usepackage{nicefrac}       
\usepackage{microtype}      

\usepackage{helvet}
\usepackage{courier}
\usepackage{multirow}
\usepackage{graphicx}
\usepackage{enumitem}
\usepackage{ragged2e}
\usepackage{epsfig}
\usepackage{dsfont}
\hypersetup{
    colorlinks=true,
    citecolor=blue, 
    linkcolor=blue,
    filecolor=magenta,      
    urlcolor=blue,
linktocpage}

\theoremstyle{plain}

\newtheorem{theorem}{Theorem}
\newtheorem{lemma}{Lemma}

\newtheorem{assumption}{Assumption}
\newtheorem{definition}{Definition}
\newtheorem{remark}{Remark}

\usepackage{algorithm}
\usepackage{algorithmic}
\renewcommand{\algorithmicinput}{\textbf{Input:}}
\renewcommand{\algorithmicoutput}{\textbf{Output:}}

\newlength{\commentindent}
\makeatletter
\renewcommand{\algorithmiccomment}[1]{\unskip\hfill\makebox[\commentindent][r]{$\rhd$~#1}\par}
\LetLtxMacro{\oldalgorithmic}{\algorithmic}
\renewcommand{\algorithmic}[1][0]{%
\oldalgorithmic[#1]%
  \renewcommand{\ALC@com}[1]{%
\ifnum\pdfstrcmp{##1}{default}=0\else\algorithmiccomment{##1}\fi}%
}

\title{A Game-Theoretic Approach to Multi-Agent Trust Region Optimization}

\author{%
 Ying Wen$^1$\thanks{Correspondence to Ying Wen <ying.wen@sjtu.edu.cn>.}, Hui Chen$^2$, Yaodong Yang$^2$, Zheng Tian$^2$, Minne Li$^2$, Xu Chen$^3$ and Jun Wang$^2$\\
 $^1$Shanghai Jiao Tong University, $^2$University College London, $^3$Renmin University.
}

\begin{document}
\maketitle

\begin{abstract}
Trust region methods are widely applied in single-agent reinforcement learning problems due to their monotonic performance-improvement guarantee at every iteration.
Nonetheless, when applied in multi-agent settings, the guarantee of trust region methods no longer holds because an agent's payoff is also affected by other agents' adaptive behaviors.
To tackle this problem, we conduct a game-theoretical analysis in the policy space, and propose a multi-agent trust region learning method (MATRL), which enables trust region optimization for multi-agent learning.
Specifically, MATRL finds a stable improvement direction that is guided by the solution concept of Nash equilibrium at the meta-game level. 
We derive the monotonic improvement guarantee in multi-agent settings and show the local convergence of MATRL to stable fixed points in differential games.
To test our method, we evaluate MATRL in both discrete and continuous multiplayer general-sum games including checker and switch grid worlds, multi-agent MuJoCo, and Atari games. Results suggest that MATRL significantly outperforms strong multi-agent reinforcement learning baselines.
\end{abstract}


\section{Introduction}
Multi-agent systems (MASs)~\cite{shoham2008multiagent} have received much attention from the reinforcement learning community \cite{yang2020overview}. In the real world, automated driving~\cite{cao2012overview,zhou2020smarts}, StarCraft II~\cite{vinyals2019grandmaster,peng2017multiagent} and Dota 2~\cite{berner2019dota} are a few examples of the myriad of applications that can be modeled by MASs.
Due to the complexity of multi-agent problems~\cite{chatterjee2004nash}, an investigation into whether agents can learn to behave effectively during interactions with environments and other agents is essential~\cite{fudenberg1998theory}.
This investigation can be conducted naively through an \emph{independent learner} (IL)~\cite{tan1993multi}, which ignores the other agents and optimizes the policy assuming a stable environment~\cite{Busoniu2010multiagentRL,hernandez2017survey}; and \emph{trust region} method (e.g., proximal policy optimization (PPO)~\cite{ppo}) based ILs are popular~\cite{vinyals2019grandmaster,berner2019dota} due to their theoretical guarantee for single-agent learning~\cite{Kakade2002ApproximatelyOA} and good empirical performance in real-world applications.

\begin{figure*}[t!]
\vspace{-10pt}
  \centering
  \includegraphics[width=0.85\textwidth]{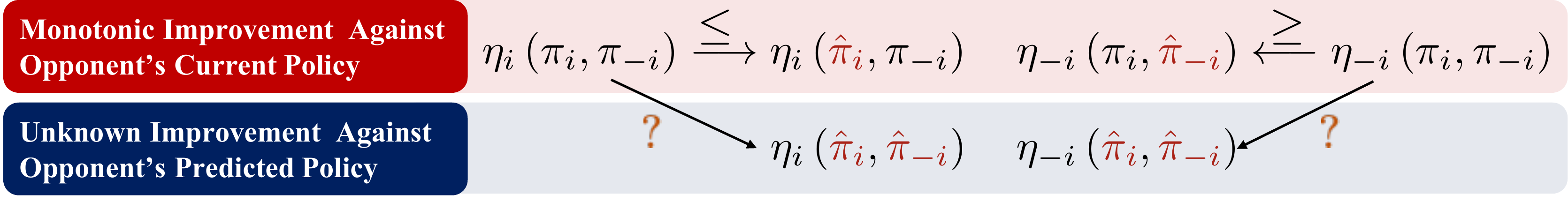}
\caption{Discounted returns $\eta_i$ for an agent $i$ given different joint policy pairs, where $\pi_i$ is the current policy, and $\pi_i^{\prime}$ is the simultaneously predicted policy. Given $\pi_i$, the monotonic improvements of a fixed opponent can be easily measured: $\eta_i(\pi_i^{\prime}, \pi_{-i}) \geq \eta_i(\pi_i, \pi_{-i})$.
However, due to simultaneous learning, the improvement of $\eta_i(\pi_i^{\prime}, \pi_{-i}^{\prime})$ is unknown compared to $\eta_i(\pi_i, \pi_{-i})$.
}
  \label{fig:rel}
 \vspace{-10pt}
\end{figure*}

In multi-agent scenarios, however, an agent's improvement is affected by other agents' adaptive behaviors (i.e., the multi-agent environment is \emph{nonstationary}~\cite{hernandez2017survey}).
As a result, trust region learners can measure the policy improvements of agents' predicted policies compared to the current policies, but the improvements compared to the other agents' predicted policies are still unknown (shown in Fig.~\ref{fig:rel}).
Therefore, trust-region-based ILs perform worse in MASs than in single-agent tasks.
Moreover, the convergence to a \emph{fixed point}, such as a \emph{Nash equilibrium}~\cite{nash1950equilibrium,bowling2004existence,mazumdar2020gradient}, is a common and widely accepted solution concept for multi-agent learning.
Thus, although ILs can best respond to other agents' current policies, they lose their convergence guarantee~\cite{laurent2011world}.

One solution for addressing the convergence problem for ILs is empirical game-theoretic analysis (EGTA)~\cite{wellman2006methods}, which approximates the best response to the policies generated by ILs~\cite{lanctot2017unified,muller2019generalized}.
Although EGTA-based methods~\cite{lanctot2017unified, omidshafiei2019alpha,balduzzi2019open} establish convergence guarantees in several game classes, their computational cost is also large when empirically approximating and solving the meta-game~\cite{yang2019alpha}.
Other multi-agent learning approaches collect or approximate additional information such as communication~\cite{jakob2016communication,peng2017multiagent} and centralized joint critics~\cite{maddpg,coma1,yang2020multi,qmix}.
Nevertheless, these methods usually require centralized critics or centralized communication assumptions, which require extra training efforts.
Thus, there is considerable interest in the use of multi-agent learning to find an algorithm that while having minimal requirements and computational cost as ILs, also simultaneously improves convergence performance.

This paper presents a \emph{multi-agent trust region learning} (MATRL) algorithm that augments the trust region ILs with a meta-game analysis to improve learning stability and efficiency.
In MATRL, a trust region trial step for an agent's payoff improvement is implemented by ILs, which provide a predicted policy based on the current policy.
Then, an empirical policy-space meta-game is constructed to compare the expected advantages of the predicted policies with those of the current policies.
By solving the meta-game, MATRL finds a restricted step by aggregating the current and predicted policies using the meta-game Nash equilibrium.
Finally, MATRL takes the best responses based on the aggregated policies from the last step for each agent to explore because the identified stable trust region is not always strictly stable.
MATRL is, therefore, able to provide a weakly stable solution compared to naive ILs.
Based on a trust region IL, MATRL requires the knowledge of other agents' policy during the meta-game analysis but does not need extra centralized parameters, simulations, or modifications to the IL itself.
We provide insights into the empirical meta-game in Section~\ref{sec:meta_game}, showing that the approximated Nash equilibrium of the meta-game is a weak stable fixed point of the underlying game.
Our experiments demonstrate that MATRL significantly outperforms deep ILs~\cite{ppo} with the same hyperparameters, VDN~\cite{vdn}, QMIX~\cite{qmix} and QDPP~\cite{yang2020multi} methods in discrete action grid worlds, decentralized PPO ILs, centralized MADDPG~\cite{maddpg} and independent DDPG 
and COMIX~\cite{witt2020deep}
in a continuous action multi-agent MuJoCo task~\cite{witt2020deep} and zero-sum multi-agent Atari~\cite{terry2020arcade}.
\section{Multi-Agent Trust Region Learning}
\label{sec:matrl}

 \textbf{Notations \& Preliminaries.} A stochastic game~\cite{shapley1953stochastic,littman1994markov} can be defined as follows:
$\mathcal{G} = \langle \mathcal{N}, \mathcal{S},
\{\mathcal{A}_{i}\},
\{\mathcal{R}_{i}\},
\mathcal{P}, p_0, \gamma \rangle$,
where $\mathcal{N}$ is a set of agents, $n=|\mathcal{N}|$ is the number of agents, and $\mathcal{S}$ denotes the state space.
$\mathcal{A}_i$ is the action space for agent $i$. $\mathcal{A}=\mathcal{A}_1 \times \cdots \times \mathcal{A}_{n}=\mathcal{A}_i \times \mathcal{A}_{-i}$ is the joint action space, and for simplicity, we use $-i$ to denote agents other than agent $i$.
$\mathcal{R}_i=R_i(s, a_i, a_{-i})$ is the reward function for agent $i \in \mathcal{N}$.
$\mathcal{P}:\mathcal{S}\times\mathcal{A}\times \mathcal{S} \rightarrow [0,1]$ is the transition function.
$p_0$ is the initial state distribution, and $\gamma \in [0,1)$ is a discount factor.
Each agent $i \in \mathcal{N}$ has
a stochastic policy $\pi_i(a_i|s):\mathcal{S}\times\mathcal{A}_i\rightarrow [0,1]$
and aims to maximize its long-term discounted return:
\begin{equation}
\label{eq:rtn}
   \eta_i(\pi_i, \pi_{-i}) = \mathbb { E }_{s^0, a_i^0, a_{-i}^0 \cdots} \left[ \sum _ { t = 0 } ^ { \infty } \gamma ^ { t } R_i( s^{t}, a_i ^ { t }, a_{-i} ^ { t  }  ) \right],
\end{equation}
where $ s^0 \sim p_0$, $s^{t+1} \sim \mathcal{P}(s^{t+1} | s^t, a_i^t, a_{-i}^t)$, and $a_i^t \sim \pi_i(a_i^t|\tau_i^t)$. Then, we have the standard definitions of the state-action value and state value functions: $Q_i^{\pi_i, \pi_{-i}}(s^t, a_i^t, a_{-i}^t)=\mathbb { E }_{s^{t+1}, a_i^{t+1}, a_{-i}^{t+1} \cdots} [ \sum _ { l = 0 } ^ { \infty  } \gamma ^ { l } R_i(s^{t+l}, a_i ^{t+l}, a_{-i} ^{t+l}  )]$ and $V_i^{\pi_i, \pi_{-i}}(s^t)=\mathbb { E }_{a_i^{t}, a_{-i}^{t}, s^{t+1} \cdots} [ \sum _ { l = 0 } ^ { \infty  } \gamma ^ { l } R_i( s^{t+l}, a_i ^ { t+l }, a_{-i} ^ { t+l}  )]$; also the advantage function $A_i^{\pi_i, \pi_{-i}}(s^t, a_i^t, a_{-i}^t)=Q_i^{\pi_i, \pi_{-i}}(s^t, a_i^t, a_{-i}^t)-V_i^{\pi_i, \pi_{-i}}(s^t)$, given the state and joint action.

\begin{figure*}[t!]
  \centering
\vspace{-10pt}
  \includegraphics[width=1.\textwidth]{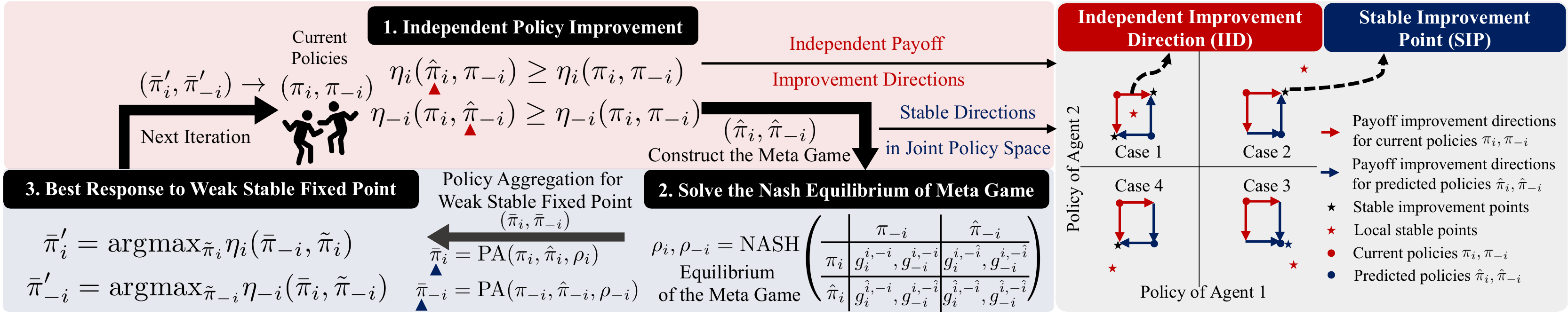}
\caption{\textbf{(Left)}: Overview of the MATRL phases. The pale red area indicates independent payoff improvement directions; the pale blue area shows stable improvement directions in joint policy space and $\pi$: current policy, $\hat{\pi}$: predicted policy in IID step, $\bar{\pi}$: aggregated policy in SIP step;  $\pi'$, next policy.
\textbf{(Right)}: the gray area illustrates IID and SIP with a two-agent game, in which the arrows indicate the payoff improvement directions for agents. The IID guarantees the partially monotone game in red arrows; then, the SIPs are determined by improvement directions (include four cases) of predicted policies in blue arrows.}
  \label{fig:trl}
 \vspace{-10pt}
\end{figure*}

\textbf{Motivations.} A trust region algorithm aims to answer two questions: how to compute a trial step and whether the trial step should be accepted.
In multi-agent learning, a trial step toward agents payoff improvement can be easily implemented with ILs, denoted as \textbf{\emph{independent improvement direction (IID)}}. 
The remaining issue is resolved by finding a restricted step leading to a stable improvement direction, which is not in the single agent's policy space but in the joint policy space.
In other words, MATRL decomposes trust region learning into two parts: first, an IID between \emph{current policy} $\pi_i$ and \emph{predicted policy} $\hat{\pi}_i$ should be identified; then, with the help of the predicted policy, a more refined method, to some extent, can approximate a stable trial step.
Instead of line searching in a single-agent payoff improvement~\cite{trpo} direction, MATRL searches for a joint policy space to achieve a conservative and stable improvement. 
Essentially, MATRL is an extension of the single-agent TRPO to a MAS, which learns to find a stable point between the current policy and the predicted policy.
To find the stable improvement directions, we assume knowledge about other agents' policies during training to avoid unstable improvement via empirical meta-game analysis, while the execution can still be fully decentralized.
We explain every step of MATRL in detail in the following sections (also in Fig.~\ref{fig:trl}).

\subsection{Independent Trust Payoff Improvement}
Single-agent reinforcement learning algorithms can be straightforwardly applied to multi-agent learning, where we assume that all agents behave independently~\cite{tan1993multi}.
In this section, we have chosen the policy-based reinforcement learning method---ILs.
In multi-agent games, the environment becomes a Markov decision process for agent $i$ when each of the other agents plays according to a fixed policy. We set agent $i$ to make a monotonic improvement against its opponents' fixed policies.
Thus, at each iteration, the policy is updated by maximizing the utility function $\eta_i$ over a local neighborhood of the current joint policy $\pi_i, \pi_{-i}$.
 We can adopt TRPO (or, PPO~\cite{ppo}), which constrains the step size in the policy update:
\begin{equation}
\label{eq:pi_hat}
  \hat{\pi}_i=\arg \max _{\pi \in \Pi_{\theta_i}} \eta_i(\pi , \pi_{-i}) \quad \mathrm{s.t.}~ D\left(\pi_i, \hat{\pi}_{i}\right) \leq \delta_i,
\end{equation}
where $D$ is a distance measurement, and $\delta_i$ is a constant. Independent trust region learners produce the monotonically improved policy $\hat{\pi}_{i}$, which guarantees $\eta_{i}\left(\hat{\pi}_{i}, \pi_{-i}\right) \geq \eta_{i}\left(\pi_{i}, \pi_{-i}\right)$ and provides a trust payoff bound by $\hat{\pi}_{i}$.
Due to simultaneous policy improvement without awareness of other agents
, however, the lower bound of payoff improvement from single-agent~\cite{trpo} no longer holds for multi-agent payoff improvement.
By following a similar logic in proof, we can obtain a precise lower bound for a simultaneous-move multi-agent payoff improvement.
\begin{remark}
\label{the:payoff_region}
The approximated expected advantage $g_{i}^{\pi_{i},\pi_{-i}}$ gained by agent $i$ when $\pi_{i}, \pi_{-i} \rightarrow \hat{\pi}_{i}, \hat{\pi}_{-i}$ is denoted as follows:
\begin{equation}
\label{eq:adv}
\begin{aligned}
g_{i}^{\pi_{i},\pi_{-i}} (\hat{\pi}_{i}, \hat{\pi}_{-i}) := 
\sum_{s}p^{\pi_i,\pi_{-i}}(s)\sum_{a_i,a_{-i}}\hat{\pi}_i(a_i|s)\hat{\pi}_{-i}(a_{-i}|s)
A_{i}^{\pi_{i}, \pi_{-i}} (s, a_{i}, a_{-i}),
\end{aligned}
\end{equation}
where $p^{\pi_i,\pi_{-i}}(s)$ discounted state visitation frequencies induced by $\pi_i,\pi_{-i}$.
Then, the following lower bound can be derived for multi-agent independent trust region optimization:
\begin{equation}
\begin{aligned}
	\eta_i(\hat{\pi}_i, \hat{\pi}_{-i})-\eta_i(\pi_i, \pi_{-i}) \geq 
	 g_i^{\pi_i, \pi_{-i}}(\hat{\pi}_i, \hat{\pi}_{-i})-\frac{4 \gamma \epsilon_i}{(1-\gamma)^2}(\alpha_i+\alpha_{-i}-\alpha_i\alpha_{-i})^2,
\end{aligned}
\end{equation}
where $\epsilon_i =\max_{s, a_{-i}, a_{-i}} \big| A_i^{\pi_i, \pi_{-i}}(s, a_i, a_{-i})\big|$, $\alpha_i=D_{\mathrm{TV}}^{\max }\left(\pi_{i}, \hat{\pi}_{i}\right)=\max _{s} D_{\mathrm{TV}}(\pi_i(\cdot | s) \| \hat{\pi}_i(\cdot | s))$ for agent $i$, and $D_{\mathrm{TV}}$ is the total variation divergence~\cite{trpo}. More details are included in Appendix~\ref{kl_trust_region_proof}.
\end{remark}

Based on the independent trust payoff improvement, although the predicted policy $\hat{\pi}_i$ will guide us in determining the step size of the IID, the stability of $(\hat{\pi}_i, \hat{\pi}_{-i})$ is still unknown.
As shown in Remark~\ref{the:payoff_region}, an agent's lower bound is approximately $O(4\alpha^2)$, which is four times larger than the single-agent lower bound trust region of $O(\alpha^2)$~\cite{Kakade2002ApproximatelyOA}.
Furthermore, $\epsilon_i =\max_{s, a_{-i}, a_{-i}} \big| A_i^{\pi_i, \pi_{-i}}(s, a_i, a_{-i})\big|$ depends on other agents' action $a_{-i}$, which will be very large when agents have conflicting interests.
Therefore, the most critical issue underlying MATRL is finding a Stable Improvement Point (SIP) after the IID.
In the next section, we illustrate how to search for a weak stable fixed point within the IID based on the  meta-game analysis.
\subsection{Approximating the Weak Stable Fixed Point}
\label{sec:meta_game}

Stabilizing the independent trust payoff improvements is one of the essential components of MATRL.
Since each iteration of MATRL requires the solving of
additional stable improvement subproblem, finding an efficient solver for this subproblem is very important.
Instead of using the stable fixed points~\cite{balduzzi2018mechanics} as the stable improvement target, we choose the \emph{weak stable fixed point} in Definition~\ref{def:wsfp}, which is much easier to find.
To maximize the objective defined in Eq.~(\ref{eq:rtn}), we can ask that \emph{reasonable} algorithms avoid all strict minimums (unstable fixed points), which imposes only that agents are well-behaved regarding strict minima, even if their individual behaviors are not self-interested~\cite{letcher2020impossibility}. Before providing the clear definitions for these points, we first define a differentiable game restricted by the IID:
\begin{definition}[Differentiable Restricted Game (DRG)]
\label{def:rug}
If the policy space for each agent $i$ in a game is restricted to open sets $\bar{\Pi}_i = [\pi_i, \hat{\pi}_i] \subseteq \Pi_i$, where $\bar{\Pi}_i \subseteq \Pi_i$, and the expected advantage $g_{i}$ is twice continuously differentiable in this range, then we call it a differentiable restricted game.
\end{definition}
Denote the simultaneous gradient of the DRG as $\boldsymbol{\xi}(\pi_i,\pi_{-i})=(\nabla_{\mathbf{\pi}_{i}} g_{i}, \nabla_{\mathbf{\pi}_{-i}} g_{-i})$. 
Adapted from \cite{balduzzi2018mechanics} and \cite{letcher2020impossibility}, we introduce the Hessian of DRG as the block matrix $H=\nabla_{\pi_i,\pi_{-i}}\boldsymbol{\xi}(\pi_i,\pi_{-i})$ to define the types of fixed points:
\begin{definition}[Weak Stable Fixed Point]
\label{def:wsfp}
A point $(\bar{\pi}_i,\bar{\pi}_{-i})$ is a fixed point if $\bs{\xi}(\bar{\pi}_i,\bar{\pi}_{-i})=\bs{0}$.
We then say that $(\bar{\pi}_i,\bar{\pi}_{-i})$ is stable if  $H(\bar{\pi}_i,\bar{\pi}_{-i})\preceq 0$, is unstable if  $H(\bar{\pi}_i,\bar{\pi}_{-i})\succ 0$ and is a \textbf{weak stable fixed point} if $H(\bar{\pi}_i,\bar{\pi}_{-i})\nsucc 0$\footnote{In this paper, we want to maximize the return, not minimize the loss, so we need to avoid a strict minimum.}.
\end{definition}

We denote the weak stable fixed points in the DRG as the \textbf{\emph{stable improvement point (SIP)}}, it is reasonable if it converges only to fixed points and avoids unstable fixed points (strict minimum) almost completely.
Given that we already have the IID, which produces a predicted policy, with the knowledge about all agents policies, it is natural to conduct an EGTA~\cite{tuyls2018generalised} to search for a SIP in the area bounded by the current and predicted policy pair.
We then define a meta-game in which each agent $i$ has only two strategies $\pi_i,\hat{\pi}_{i}$:
\begin{equation}
\label{eq:meta_game}
  \mathcal{M}(\pi_{i}, \hat{\pi}_{i}, \pi_{-i}, \hat{\pi}_{-i})=\left(\begin{array}{cc}g_{i}^{i,-i}, g_{-i}^{i,-i} & g_{i}^{i,-\hat{i}}, g_{-i}^{i,-\hat{i}} \\ g_{i}^{\hat{i},-i}, g_{-i}^{\hat{i},-i} & g_{i}^{\hat{i},-\hat{i}}, g_{-i}^{\hat{i},-\hat{i}}\end{array}\right),
\end{equation}
where $g_{i}^{\hat{i},-\hat{i}} = g_{i}^{\pi_i,\pi_{-i}}(\hat{\pi}_i,\hat{\pi}_{-i})$ (as defined in Eq.~(\ref{eq:adv})) is an empirical payoff entry of the meta-game, and note that $g_{i}^{i,-i}=0$, as it has an expected advantage over itself.
Compared with using $\eta_i(\hat{\pi}_i, \hat{\pi}_{-i})=\eta_i(\pi_i, \pi_{-i}) + g_{i}^{\hat{i},-\hat{i}}$ as the meta-game payoff, $g_{i}^{\hat{i},-\hat{i}}$ has lower variance and is easier to approximate because $\eta_i(\pi_i, \pi_{-i})$ is a constant baseline.
However, most entries in $\mathcal{M}$ are unknown, and many extra simulations are required to estimate the payoff entries (e.g., $g_{i}^{\hat{i},-\hat{i}}$) in EGTA.
Instead, we reuse the trajectories in the IID step to approximate $g_{i}^{\hat{i},-\hat{i}}$ by ignoring the small changes in the state visitation density caused by $\pi_i \rightarrow \hat{\pi}_i$.
\begin{remark}
\label{re:2}
	The meta-game $\mathcal{M}(\pi_{i}, \hat{\pi}_{i}, \pi_{-i}, \hat{\pi}_{-i})$ is a partially monotone game and has a pure strategy equilibrium~\cite{takeshita2012necessity}, because the monotonic improvements $g_{i}^{i,-i} \leq g_{i}^{\hat{i},-i}$ and $g_{-i}^{i,-i} \leq g_{-i}^{i,-\hat{i}}$ when $\pi_i,\pi_{-i} \rightarrow \hat{\pi}_i,\hat{\pi}_{-i}$.
\end{remark}
Taking the two-agent case as an example, as we can see in Eq.~(\ref{eq:meta_game}), meta-game $\mathcal{M}$ becomes a $2\times2$ matrix-form game, which is much smaller in size than the whole underlying game. Besides, according to Fig.~\ref{fig:trl} Right and Remark~\ref{re:2}, all four cases have at least one pure strategy that leads a stable improvement direction.
To this end, we can use the existing Nash solvers (e.g.,CMA-ES~\cite{hansen2003reducing}) for matrix-form games to compute a Nash equilibrium $\rho_{i}, \rho_{-i} = \mathrm{NashSolver}(\mathcal{M})$ for meta-game $\mathcal{M}$, where $\rho_{i}$ and $\rho_{-i}$ $\in [0,1]$, and the Nash equilibrium of the meta-game is also an approximated equilibrium of the restricted underlying game~\cite{tuyls2020bounds}. Then, SIP policies $\bar{\pi}_{i}, \bar{\pi}_{-i}$ can be aggregated based on current policy $\pi_i$ and predicted policy $\hat{\pi}_i$ in the IID for each agent $i$.
\begin{assumption}
\label{ass:linear}
In the IID step, ILs enjoy the monotonic improvement against fixed opponent policies, in which the change from $\pi_i$ to $\hat{\pi}_i$ is usually constrained by a small step size. 
Therefore, it is reasonable to assume that there is a linear, continuous and monotonic change in the restricted policy space between $\pi_i$ and $\hat{\pi}_i$.
\end{assumption}
In this case, with $\rho_i$ being agent $i$'s Nash equilibrium policy in the meta-game, $\bar{\pi}_{i}$ can be derived via a linear mixture:
$
\bar{\pi}_{i}=\rho_{i}\pi_{i}+(1-\rho_{i})\hat{\pi}_{i}
$, which delimits agent $i$'s SIP. Now, we can prove that $(\bar{\pi}_{i}, \bar{\pi}_{-i})$ is a weak stable fixed point for the underlying game in Theorem~\ref{def:meta_nash}. 
Furthermore, based on Assumption~\ref{ass:linear}, the payoff and policy space $[\pi_i, \hat{\pi}_i]$ for DRG are bounded in a linear continuous space, we can conclude the following theorem:	
\begin{theorem}[Existence of a Weak Stable Fixed Point]
\label{def:meta_nash}
 If $(\rho_i, \rho_{-i})$ is a Nash equilibrium of the meta-game $\mathcal{M}$, then linear mixture joint policy $(\bar{\pi}_i, \bar{\pi}_{-i})$ is a weak stable fixed point for the DRG.
\begin{proof}
\vspace{-10pt}
See Appendix~\ref{app:stable_point}.
\vspace{-5pt}
\end{proof}
\end{theorem}

According to Theorem~\ref{def:meta_nash}, $(\bar{\pi}_i, \bar{\pi}_{-i})$ is a weak stable fixed point of the restricted underlying game.
Although the weak stable fixed point is relatively weak compared to the stable fixed points~\cite{balduzzi2018mechanics}, as we have stated, a weak stable fixed point is a reasonable (not as strong as it is rational) requirement for an algorithm to avoid the minimum.
Furthermore, weak stable fixed points can suit general game settings. As shown in Appendix~\ref{app:stable_point}, in cooperative, competitive, and general-sum games, the fixed points found by the meta-game analysis can be either stable or saddle points.
Similarly, a local Nash equilibrium can be stable or saddle in different games~\cite{mazumdar2020gradient}. Therefore, the goodness of stable concepts depends on specific settings.
If we make some additional game class assumptions, then we can easily obtain stronger fixed point types. Nevertheless, this approach comes with a cost, requiring additional computation or assumptions that may break the most general settings.
In addition, when the meta-game has multiple Nash equilibria, an equilibrium is randomly selected in our work.
\begin{wrapfigure}{r}{0.58\textwidth}
\vspace{-20pt}
\begin{minipage}[b]{0.58\textwidth}
\begin{algorithm}[H]
\begin{algorithmic}[1]
\INPUT Initialize policies $\pi_i$ for each $i$.
\FOR{$k \in \{0,1,2,\cdots\}$}
      \STATE Use current policies $\pi_i$, $\pi_{-i}$ to collect trajectories.
\FOR{\textbf{each} $i$}
        \STATE Compute a one-step predicted policy $\hat{\pi}_i$ (Eq.~(\ref{eq:pi_hat})).
\ENDFOR
        \STATE Solve meta-game $\mathcal{M}(\pi_{i}, \hat{\pi}_{i}, \pi_{-i}, \hat{\pi}_{-i})$.
    \STATE Compute weak stable fixed point $\bar{\pi}_{i}$, $\bar{\pi}_{-i}$.
    \STATE \textbf{For each} $i$: Compute best response $\pi_i^{\prime}$ (Eq.~(\ref{eq:br})).
     \STATE $\pi_i \leftarrow \pi_i^{\prime}$, $\pi_{-i} \leftarrow \pi_{-i}^{\prime}$.
\ENDFOR
\end{algorithmic}
\captionsetup{type=algorithm}
\caption{MATRL}
\label{algo:main_algo1}
\end{algorithm}
\end{minipage}
\vspace{-5pt}
\end{wrapfigure}

\subsection{Improvement over a Weak Stable Fixed Point}

Although the weak stable fixed point, $(\bar{\pi}_i, \bar{\pi}_{-i})$, binds the policy update to another fixed point, there are still fully stable points according to Theorem~\ref{def:meta_nash}.
Besides, it is difficult to generalize for the other parts of the policy space not reached by SIP, especially in anticoordination games~\cite{lanctot2017unified}.
Similar to the extragradient method~\cite{mertikopoulos2018optimistic}, to encourage the exploration, we apply the best response against the weak stable fixed point $(\bar{\pi}_{i}, \bar{\pi}_{-i})$:
\begin{equation}
\label{eq:br}
\pi_{i}^{\prime} =\arg \max _{\pi \in \Pi_{\theta_i}} \eta_{i}\left(\pi, \bar{\pi}_{-i}\right).
\end{equation}
To perform the best response, we need another round to collect the experiences and perform a gradient step in Eq.~(\ref{eq:br}).
However, in practice, since we already have the trajectories in the IID step, the best response to the weak stable fixed point can be easily estimated through importance sampling.
Alternatively, by defining $c_{i} \stackrel{\text { def }}{=}\min\Big(1+\bar{c}, \max (1-\bar{c}, \frac{\pi_i(a_{i} | s)}{\bar{\pi}_i(a_{i} | s)})\Big)$ as truncated importance sampling weights, we can rewrite the best response update to Eq.~(\ref{eq:br}) as an equivalent form to the following one in terms of expectations: $\pi_i^{\prime} = \arg \max _{\pi} \mathbb{E}_{a_{-i} \sim \bar{\pi}_{-i}}[c_{-i} \eta_{i}\left(\pi_{i}, \pi_{-i}\right)]$. If the agents end up playing the BR, then there is no further improvement in the IID step; the payoff entries in the restricted meta-game would be zero, meaning agents will stay at the current policies following MATRL steps.

\subsection{Local Convergence}
\begin{wrapfigure}{r}{0.4\textwidth}
\begin{minipage}[b]{0.4\textwidth}
  \centering
  \includegraphics[width=.99\linewidth]{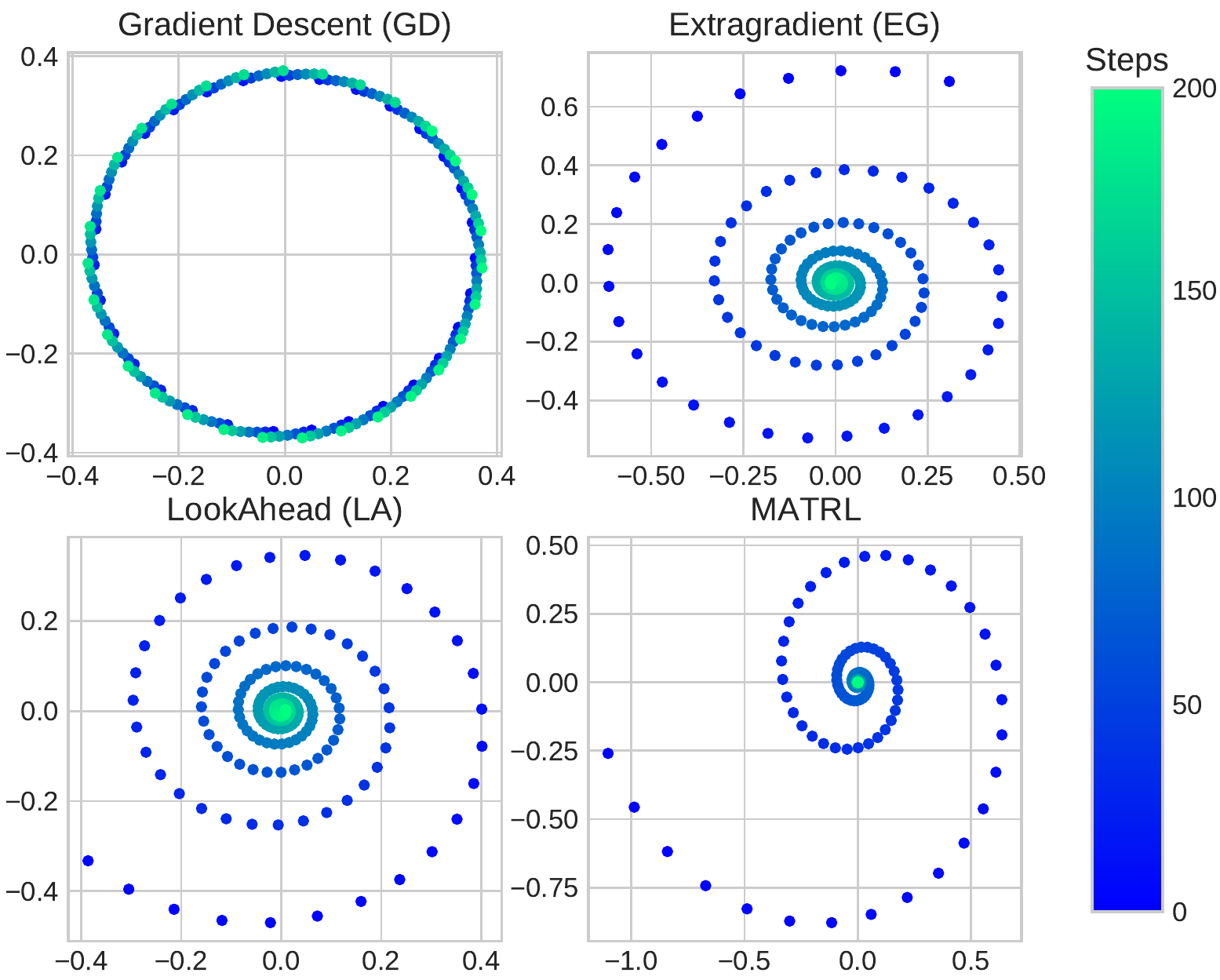}
  \captionsetup{type=figure}
\caption{Learning the dynamics of MATRL in a rotational differential game. 
GD cannot converge; EG, LA and MATRL converge to the stable fixed point and MATRL has the fastest convergence speed with same learning rate $0.02$.
}
\label{fig:diff}
 \end{minipage}
 \vspace{-10pt}
\end{wrapfigure}
MATRL is a gradient-based algorithm with the best response to policies within the SIPs, which is essentially a variant of LookAhead methods~\cite{zhang2010multi,lola,letcher2018stable}.
More specifically, MATRL enhances the classic LookAhead method with variable step size scaling~\cite{bowling2002multiagent} or two time-scale update rules~\cite{heusel2017gans} at each SIP step, which is controlled by restricted meta-game analysis.
It has been proven that the LookAhead method can locally converge to a stable fixed point and avoid strict saddles in all differentiable games~\cite{letcher2018stable,daskalakis2020ind, zhang2019conv}. 
Similarly, we show the local convergence of MATRL in Theorem~\ref{the:conv}.
Please note, here, that to investigate the convergence, fixed point iterations are conducted on the whole learning process, while the meta-game analysis step in MATRL borrows the variable stepsize scaling and shows it is reasonable to locally avoid unstable fixed points.
Unlike LOLA, which uses a first-order Taylor expansion to estimate the best response to a predicted policy, we elaborately design the look-ahead step within the SIPs and perform the gradient steps for the best response to the SIPs. We also show that MATRL empirically outperforms the typical LookAhead method, IL LookAhead (IL-LA), in the experiments. As shown in Fig.~\ref{fig:diff}, we compare the convergence of gradient decent IL, Extragradient, LookAhead and MATRL in toy differential game\footnote{A two-agent differential game adopted from~\cite{balduzzi2018mechanics}. The loss functions: $\eta_{i}(\pi_{i}, \pi_{-i})=\frac{1}{2} \pi_{i}^{2}+10 \pi_{i}\pi_{-i}, \eta_{-i}(\pi_{i}, \pi_{-i})=\frac{1}{2} \pi_{-i}^{2}-10 \pi_{i}\pi_{-i}$.} with strong rotational force, where MATRL has faster convergence to the  stable fixed point.

%
%

\begin{theorem}[Local Convergence of MATRL]
\label{the:conv}
Let the objectives $\eta_{i}(\pi_i, \pi_{-i})$ of agents are twice continuously differentiable and step size $\alpha$ is sufficiently small, MATRL converges locally to a stable fixed point with $\epsilon$ error in Euclidean distance.
\begin{proof}
\vspace{-10pt}
See Appendix~\ref{app:the2_proof}.
\vspace{-5pt}
\end{proof}
\end{theorem}

\subsection{Discussions}
\label{disc}

\textbf{Computation Cost}. Compared to pure ILs, there are two extra cost sources in common meta-game analysis: approximating and solving the meta-game~\cite{muller2019generalized}.
In our case, the meta-game is restricted to a local two-action game, where two actions, $\pi_i$ and $\hat{\pi}_i$, are close to each other.
Reusing the IID trajectories will some estimation errors~\cite{tuyls2020bounds}, but this issue can be eased by large batch size. Then, we can enjoy this proximity property and reduce the meta-game approximation cost (without extra sampling) by reusing the collected trajectories in the IID step. 
The next crucial problem is how to solve the $n$-agent two-action meta-game, which consists of the $2^{n}$ entries of each of the $n$ payoff matrices.
Solving this meta-game is much simpler than solving the whole underlying game, which increases exponentially with state size, action size, agent number, and time horizons.
As the general-sum matrix-form game has no fully polynomial time approximation for computing Nash equilibria~\cite{chen2006computing}, it usually costs a great deal to solve the game~\cite{daskalakis2009complexity}.
However, as shown in Remark~\ref{re:2}, there always exists at least one pure Nash equilibrium in the meta-game, which can be computed in polynomial time~\cite{fabrikant2004complexity}.
Therefore, if we only require an approximated Nash equilibrium, then when $n$ is small, for example, $n\leq5$, it is affordable to find a meta-game Nash equilibrium with subexponential complexity~\cite{lipton2003playing}.
But this problem still exists when $n$ is large.
In this case, we can try a mean field approximation~\cite{yang2018mean} or utilize special payoff structure assumptions (e.g., graphical game~\cite{littman2002efficient, daskalakis2009complexity}) in the meta-game to reduce computational complexity.

\textbf{Connections to Existing Methods}. MATRL generalizes many existing methods with the best response.
In extreme cases, where the meta-game Nash equilibrium is $(\rho_i,\rho_{-i})=(1,1)$, which means that the Nash aggregated policies always maintain the current policies, MATRL degenerates to ILs. Here, we always best respond to other agents' current policy $\pi_i$ and $\pi_i^{\prime} = \arg \max _{\pi_{i}} \eta_{i}(\pi_{i}, \pi_{-i})$ following Eq.~(\ref{eq:br}).
The LookAhead~\cite{zhang2010multi,lola,letcher2018stable}, extragradient~\cite{antipin2003extragradient} and exploitability descent~\cite{spar,lockhart2019exploitabilitydescent} methods are also special instances of MATRL when meta-game Nash is $(\rho_i,\rho_{-i})=(0,0)$, which means that the best response to the most aggressive predicted policy $\hat{\pi}_{-i}$ and $\pi_i^{\prime} = \arg \max _{\pi_{i}} \eta_{i}(\pi_{i}, \hat{\pi}_{-i})$. 
More specifically, let $\xi$ denotes the game's simultaneous gradient, $H_{o}$ is the matrix of anti-diagonal blocks of $H$ (Hessian of the game), and $\alpha$ is step-size. Then we can have the updating gradient for LookAhead methods as $\left(I-\alpha H_{o}\right) \xi$. Similarly, for MATRL, we have the updating gradient $\left(I-\rho\alpha H_{o}\right) \xi$, where $\rho$ is a ratio determined by meta-game Nash to dynamically adjust the step-size at each iteration.

In summary, independent trust region learners' learning in MATRL will be constrained by a weak stable fixed point.
By analyzing the relatively simpler meta-game, we can easily approximate this weak stable fixed point without extra rollouts or simulation.
Although MATRL's training is centralized, its execution is fully decentralized, and it also does not require any extra centralized parameters or higher-order gradient computation. Fig.~\ref{fig:trl} presents an overview of MATRL. We also give the pseudocode of MATRL in Algo.~\ref{algo:main_algo1}, which is compatible with any policy-based IL.



\vspace{-5pt}
\section{Related Work}
\vspace{-5pt}
The study of gradient-based methods in multi-agent learning is quite extensive~\cite{mazumdar2020gradient,Busoniu2010multiagentRL}.
Some works on learning in games have mostly focused on adjusting the step size, which attempts to use a multitimescale learning scheme~\cite{leslie2005individual,leslie2003convergent,bowling2002multiagent} to achieve convergence.
\cite{balduzzi2018mechanics,mazumdar2019finding,letcher2018stable}
tried to utilize second-order methods to shape the step size.
However, the computational cost for second-order methods is very limiting in many cases.
Other approaches include recursive reasoning techniques \cite{wen2018probabilistic,wen2019modelling}
where agents explicitly take into account how their behaviors are going to affect their opponents during the gradient updates. 
Alternatively, MATRL approximates the second-order fixed-point information via a small meta-game with less cost compared to real Hessian computation.
An alternative augments the gradient-based algorithms with the best response to the predicted polices~\cite{antipin2003extragradient, zhang2010multi, lin2020finite,lola,spar,lockhart2019exploitabilitydescent}, which targets the challenge of instability caused by agents' change policies.
Instead of taking the best response to the approximated opponent's policy, MATRL exploits the ideas from both streams and introduces an improvement over the weak stable fixed point.


The research also focuses on the EGTA~\cite{tuyls2018generalised,jordan2009generalization,tuyls2020bounds},
which creates a policy-space meta-game for modeling multi-agent interactions.
Using various evaluation metrics, this work then updates and extends the policies based on the analysis of meta policies~\cite{lanctot2017unified, muller2019generalized,omidshafiei2019alpha,balduzzi2019open,yang2019alpha}.
Although these methods are broad with respect to multi-agent tasks, they require extensive computing resources to estimate the empirical meta-game and solve it with its increasing size~\cite{omidshafiei2019alpha,yang2019alpha}.
In our method, we adopt the idea of a policy-space meta-game to approximate the fixed point.
Unlike previous works, we only maintain current and predicted policies to construct the meta-game, which is computationally achievable in most cases.
The payoff entry in MATRL's meta-game is the expected advantage, which has a lower estimation variance compared to the commonly used empirically estimated return in EGTAs.
Regardless, we can reuse the trajectories in the IID step to estimate the payoffs without incurring additional sampling costs.

Recently, due to the use of neural networks as a function approximation for policies and values, many works have emerged on deep reinforcement learning (DRL)~\cite{dqn,ddpg}.
TRPO~\cite{Kakade2002ApproximatelyOA,trpo,ppo} is one of the most successful DRL methods in the single-agent setting, which places constraints on the step size of policy updates, monotonically preserving any improvements.
Based on the monotonic improvement in single-agent TRPO~\cite{trpo}, MATRL extends the improvement guarantee to the multi-agent level towards a weak stable fixed point.
Some works have directly applied fully decentralized single-agent DRL methods~\cite{tan1993multi}, which can be unstable during learning due to the issue of nonstationarity.
However, ~\cite{jakob2016communication,sukhbaatar2016learning,bicnet} added an extra communication channel during the training and execution in a centralized way to avoid this nonstationarity issue.
\cite{vdn, qmix, coma1, maddpg} further exploit the setting of centralized learning decentralized execution (CTDE).
These methods provide solutions for training agents in complex multi-agent environments, and the experimental results show their effectiveness compared with ILs.
Similar to the CTDE setting, MATRL also enjoys fully decentralized execution. Although MATRL still needs knowledge about other agents' policies in adjusting the step size during training, it does not need  centralized critics or any communication channels. Besides, \cite{li2021dealing, li2020multi} attempted to apply trust-region methods in networked multi-agent settings by conducting consensus optimization with their neighbors. Instead takes a game-theoretical approach to compute the meta-game Nash to find policy improvement directions without networked assumption.

\begin{figure*}[t!]
     \centering
\vspace{-15pt}
\begin{subfigure}[l]{.32\textwidth}
         \centering
         \includegraphics[width=1.\textwidth]{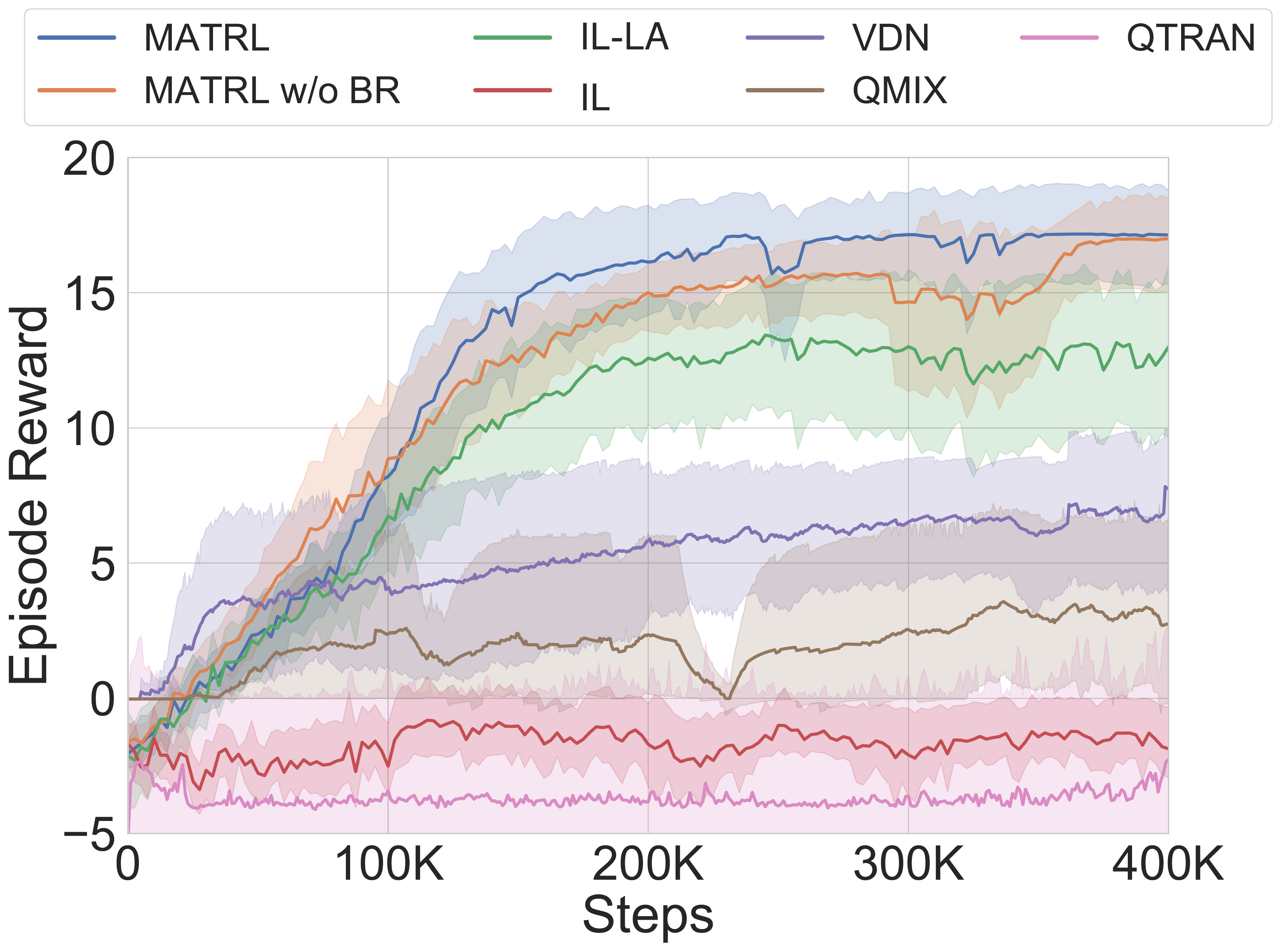}
\caption{Two-agent checker.}
         \label{fig:checker_p1}
\end{subfigure}
\begin{subfigure}[l]{.32\textwidth}
         \centering
         \includegraphics[width=1.\textwidth]{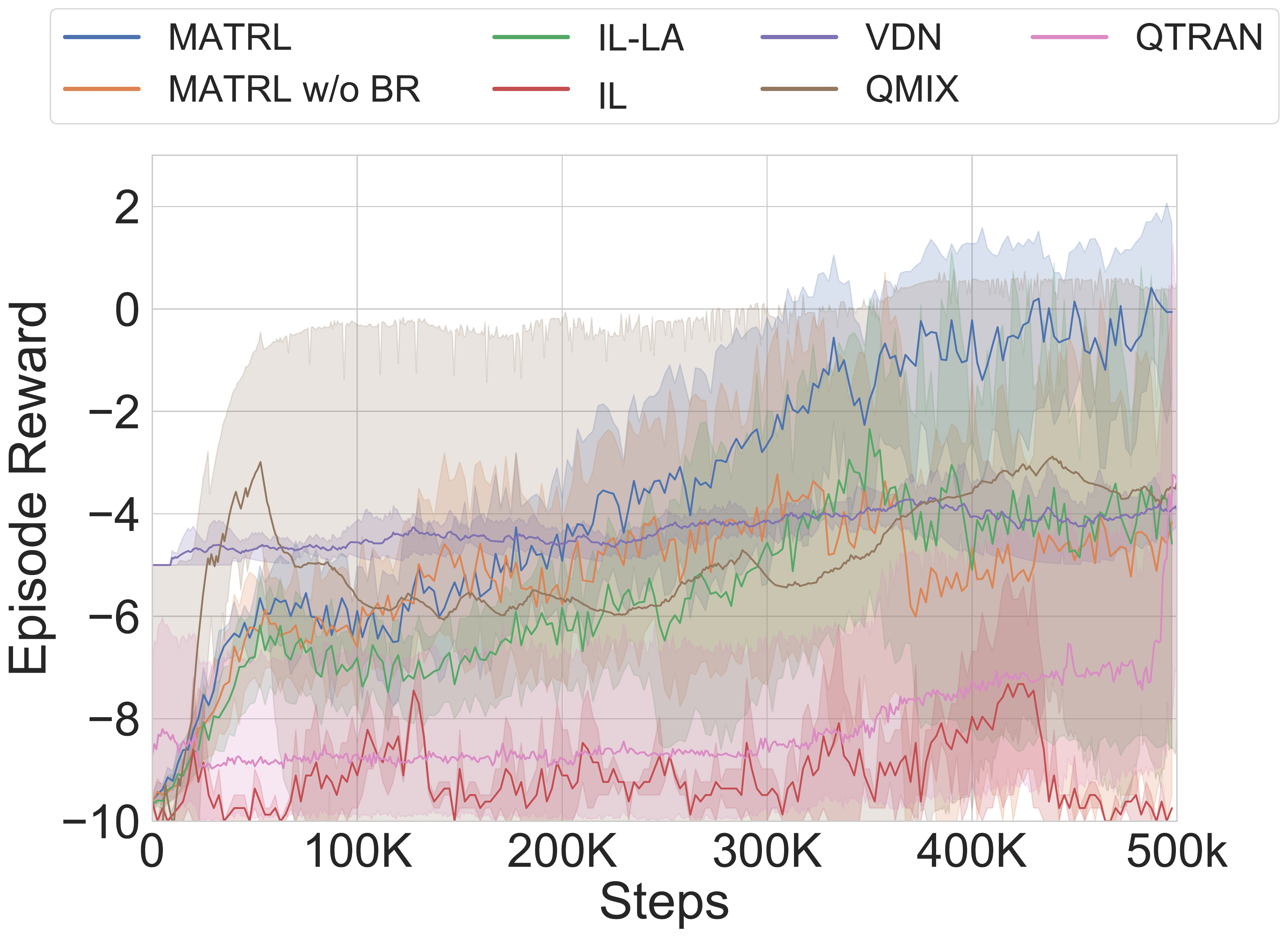}
\caption{Four-agent switch.}
       \label{fig:switch4_p1}
\end{subfigure}
\begin{subfigure}[l]{.315\textwidth}
          \centering
          \includegraphics[width=1.\textwidth]{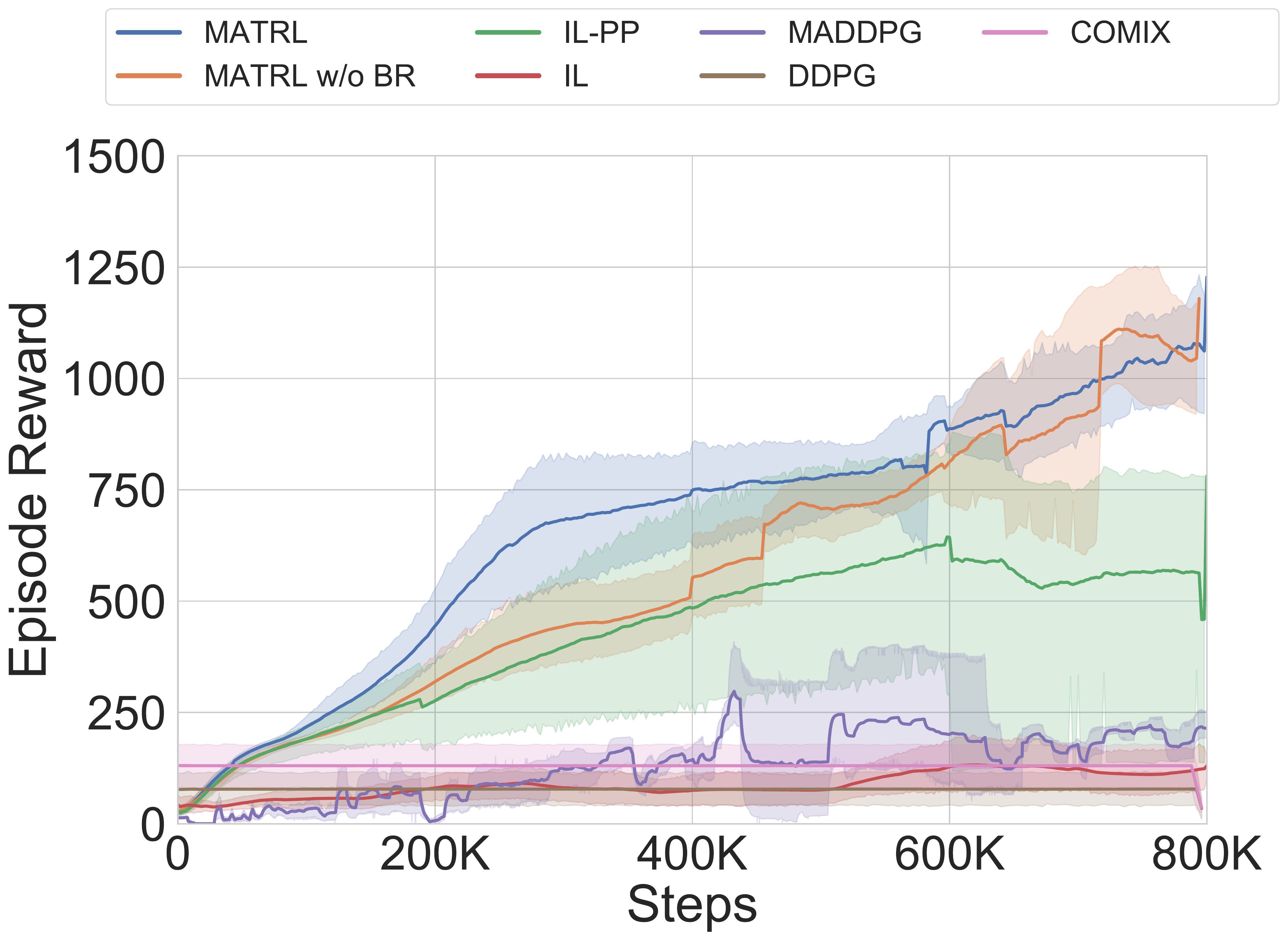}
\caption{Three-agent MuJoCo hopper.}
          \label{fig:hopper}
\end{subfigure}
 \vspace{-5pt}
\caption{Learning curves in discrete and continuous tasks. The solid lines are average episode returns with 10 random seeds for each model, and the light color areas are the error bars.}
  \label{fig:grid_learning_curve}
 \vspace{-15pt}
\end{figure*}

\vspace{-5pt}
\section{Experiments}
 \vspace{-5pt}
We design experiments to answer the following questions:
1) Can the MATRL method empirically contribute to convergence in general game settings, including cooperative/competitive and continuous/discrete games?
2) How is the performance of MATRL compared to ILs with the same hyperparameters and other strong MARL baselines in discrete and continuous games with various agent numbers?
3) Do the meta-game and best response to the weak stable fixed point bring about benefits?
We first evaluate the convergence performance of MATRL in matrix form games to answer the first question and validate the effectiveness of convergence.
For Question 2, we show that MATRL largely outperforms ILs (PPO~\cite{ppo}) and other centralized baselines (QMIX~\cite{qmix}, QTRAN~\cite{qtran}
and VDN~\cite{vdn}) in discrete grid world games that have coordination problems.
MATRL also outperforms DDPG~\cite{ddpg}, MADDPG~\cite{maddpg} and COMIX~\cite{witt2020deep} for continuous multi-agent MuJoCo games.
In addition, we test the algorithms with a 2-agent Atari Pong game to investigate whether MATRL can mitigate unstable cyclic behaviors~\cite{balduzzi2019open} in zero-sum games.
In these tasks, MATRL uses the same PPO configurations as ILs to examine the effectiveness of the trust region gradient-update mechanism,
and we use official implementations for the other baselines.
The step-by-step PPO-based MATRL algorithm is given in Appendix~\ref{app:matrl_ppo}.
Finally, ablation studies are conducted by: 1. removing the best response, called the ``MATRL w/o BR''; 2. skipping the SIP estimation, named ``IL-LA'', which has similar procedures as those of LOLA~\cite{lola,zhang2010multi}, which approximates the best response to the predicted policies via Taylor expansion, but IL-LA takes the best response gradient steps for the predicted policies.
These configurations provide insights into how much, if at all, the SIP and the best response contribute to the MATRL's performance.
We also provide more environmental details and extra experimental results, in Appendices~\ref{app:exp_env_detail} and~\ref{app:exp_detail}, with detailed experimental settings and hyperparameters used for the algorithms.
The code and experiment scripts are also anonymously available at \url{https://github.com/matrl-project/matrl}.

\begin{wrapfigure}{r}{0.75\textwidth}
\vspace{-10pt}
\begin{minipage}[b]{0.75\textwidth}
\centering
\captionsetup{type=table}
\caption{Convergence rate and average convergence step in $1,000$ random $2\times2$ matrix games. 
    MATRL shows slightly better convergence rate and speed compared to IGA-LA. }
\label{tb:matrix_game_stats}
\begin{sc}
\resizebox{1. \linewidth}{!}{
  \begin{tabular}{ llll }
  \toprule
   &\multicolumn{3}{c}{Convergence Rate (in \%) / Average Convergence Step} \\
  \cmidrule{2-4}
  \textbf{Algo.} & Coordination  & Anticoordination  & Cyclic\\
   \midrule
   \textbf{IGA} & \textbf{99} $\pm$ 0.1 / 140.67 $\pm$ 105  & 97.5 $\pm$ 0.13 / 88.95  $\pm$ 130  & 78.0 $\pm$ 0.45 / 452.92 $\pm$  202\\ 
   \textbf{IGA-LA}& \textbf{99} $\pm$ 0.1 / 138.56 $\pm$ 105 & 97.5 $\pm$ 0.08 / 83.11 $\pm$ 129  & 80.9 $\pm$ 0.43  / 432.98 $\pm$ 206 \\ 
  \hline \textbf{MATRL} & \textbf{99} $\pm$ 0.1 /  \textbf{86.54} $\pm$ 77 & \textbf{98.3 } $\pm$ 0.08 / \textbf{75.52} $\pm$ 119 & \textbf{84.6} $\pm$ 0.36  / \textbf{369.40} $\pm$ 200\\ 
  \bottomrule
  \end{tabular}
  }
\end{sc}  
\end{minipage} 
\vspace{-10pt}
\end{wrapfigure}

\textbf{Random $\mathbf{2 \times 2}$ Matrix Games}.
To adequately examine MATRL in matrix games, we randomly generate three thousand $2 \times 2$ games of three types: coordination, anticoordination, and cyclic~\cite{pangallo2017taxonomy}.
We choose the IGA and IGA-LA~\cite{zhang2010multi} as baselines and use IGA~\cite{singh2000nash} as the ILs of MATRL. The results in Table~\ref{tb:matrix_game_stats} show that MATRL has a higher convergence rate, fewer steps for convergence and more stable performance in all types of games. 
More details about game generation and the effects of the learning dynamics are provided in Appendix~\ref{app:exp_env_detail}.

\textbf{Grid Worlds}.
We evaluated MATRL in two grid world games from MA-Gym~\cite{magym}, two-agent checker, and four-agent switch, which are similar to the games  in~\cite{vdn} but with more agents to examine if MATRL can handle the games that have more than two agents.
In the checker game, two agents cooperate in collecting fruit on the map; the sensitive agent obtains $5$ for an apple and $-5$ for a lemon, while the other agent obtains $1$ and $-1$, respectively.
Therefore, the optimal solution is to let the sensitive agent obtain the apple and the less sensitive agent obtain the lemon.
In the four-agent switch game, two rooms are connected by a corridor, each room has two agents, and the four agents try to go through one corridor to the target in the opposite room.
Only one agent can pass through the corridor at one time, and agents obtain $-0.1$ for each step and $5$ for reaching the target, so they need to cooperate to obtain optimal scores.
In both games, agents can move in four directions and only partially observe their position. Although our formulation uses a fully observable setting, in this game, the methods are adapted to the partially observable setting by pretending the observation is a state.
We compare the MATRL with the PPO-based IL and two off-policy centralized training and decentralized execution baselines: VDN~\cite{vdn}, QTRAN~\cite{qtran} and QMIX~\cite{qmix}.
The results are given in Figs.~\ref{fig:checker_p1} and~\ref{fig:switch4_p1}, where MATRL shows stable improvement and outperforms other baselines.
In a two-agent checker game, using the best response, our method can achieve a total reward of $18$, while the ILs' reward stays at $-2$.
In addition, although PPO-based MATRL uses on-policy learning, it achieves better final results in fewer time steps compared to the off-policy baselines.
For the four-agent switch game, as shown in Fig.~\ref{fig:switch4_p1}, MATRL can continuously improve the total rewards to $6.5$, which is the closest to the optimal score for this game when compared with other baselines.
The result of the four-agent switch also demonstrates the effectiveness of MATRL in guaranteeing stable policy improvement for games that have more than two agents.

\begin{figure*}[t!]
\label{exp:22}
  \centering
\vspace{-15pt}
\begin{subfigure}[l]{.64\textwidth}
\includegraphics[width=1.\textwidth]{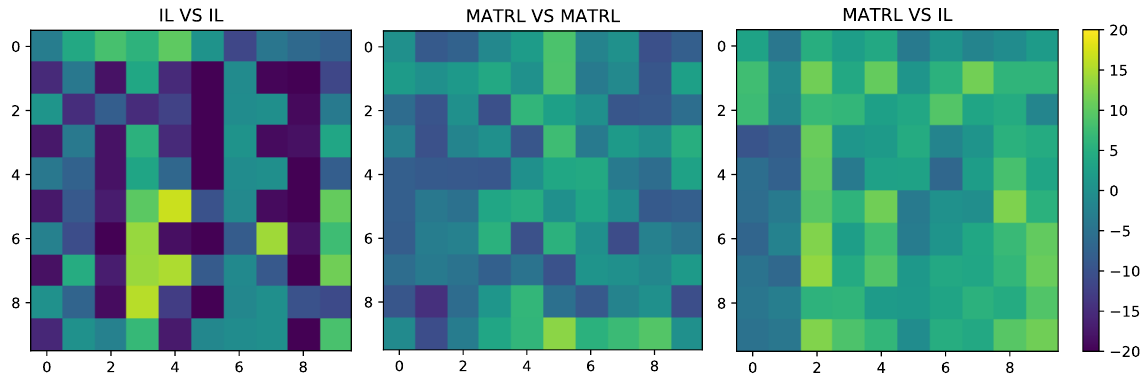}
 \vspace{-10pt}
\caption{}
  \label{fig:zerosum_all}
\end{subfigure}
\begin{subfigure}[l]{.35\textwidth}
	\includegraphics[width=1.\textwidth, height=.635\textwidth]{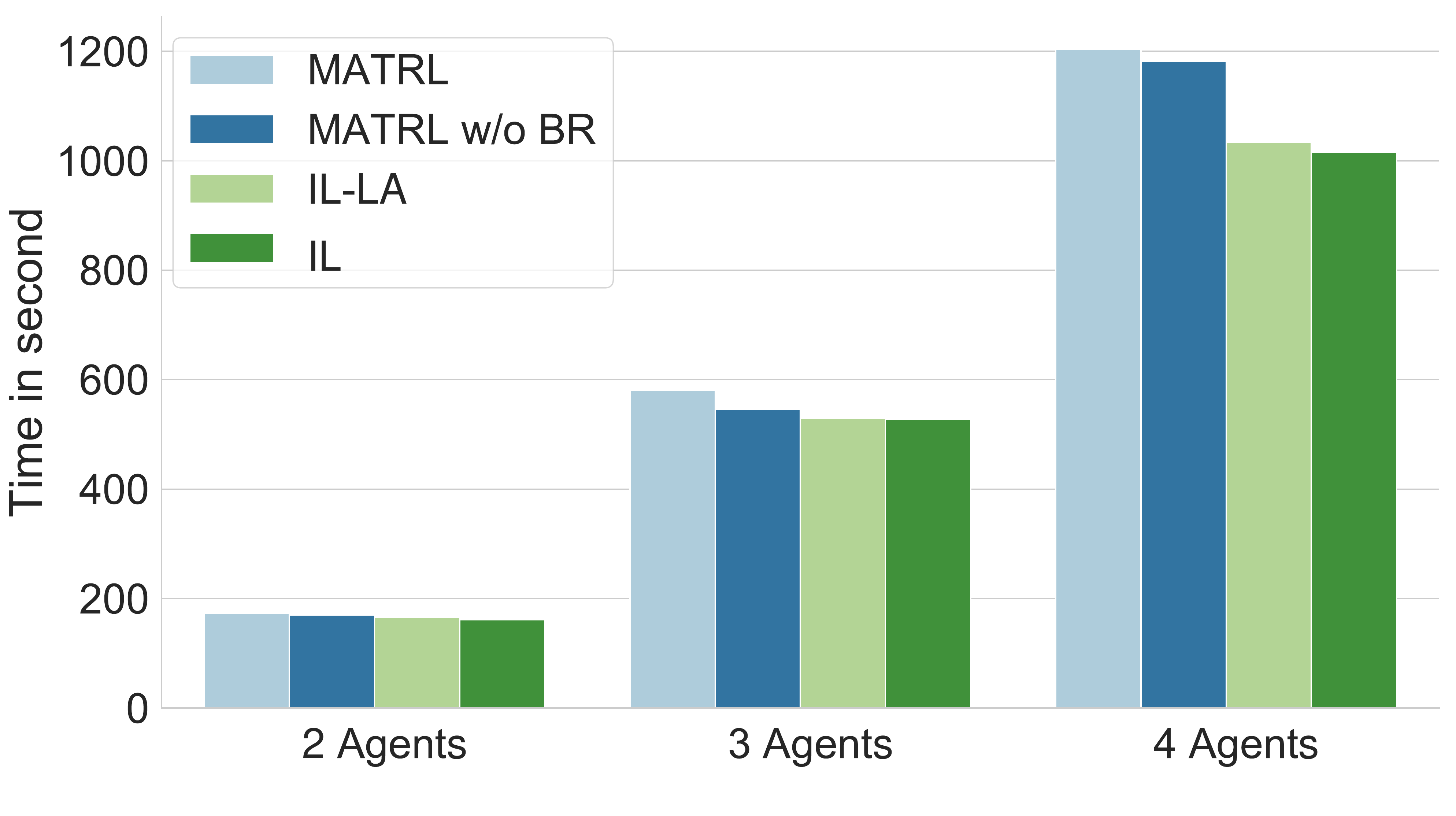}
  \vspace{-20pt}
\caption{}
  \label{fig:run_time}
\end{subfigure}
 \vspace{-5pt}
\caption{\textbf{(a)}: MATRL/IL versus MATRL/IL in the two-agent Pong game. For each setting, the grids show pairwise performance (average scores) by pitting their ten checkpoints against one another; yellow indicates a higher score. \textbf{(b)}: Run time for 20,000 environment steps (including 50 gradient steps) for the algorithms in two- to four-agent games.}
\vspace{-15pt}
\end{figure*}

\textbf{Multi-Agent MuJoCo}.
We also examined MATRL in a multi-agent continuous control task with a three-agent hopper from~\cite{witt2020deep}.
Here, three agents cooperatively control each part of a hopper to move forward.
The agents are rewarded with the distance traveled and the number of steps they make before falling.
Fig.~\ref{fig:hopper} shows that MATRL significantly outperforms ILs, MADDPG, DDPG, and the benchmarks like COMIX in~\cite{witt2020deep} within the same amount of time. More results in multi-agent MuJoCo tasks (2-agent ant and 2-agent swimmer) are available in Appendix~\ref{app:exp_env_detail}. 

\textbf{Multi-Agent Atari Pong Game}.
In the 2-agent Pong game experiments, we used raw pixels as observations and trained the MATRL and IL agents independently. Following training, we compare the pairwise performance of these models by pitting their ten checkpoints against one another and recording average scores. We report the results in Fig.~\ref{fig:zerosum_all}, which shows that MATRL outperforms ILs in MATRL vs. IL settings in most policy pairs. In addition, from the MATRL vs. MATRL and ILs vs. IL settings' results, we can see that MATRL has a more transitive learning process than that of ILs, which means that MATRL can mitigate the common cyclic behaviors in zero-sum games.
\textbf{Effect and Cost of the SIP and Best Response to a Fixed Point}.
This section analyzes the effect of the SIP from the meta-game Nash equilibrium and the best response against the weak stable fixed point.
The ablation settings are obtained by removing the SIP (IL-LA) and the best response (MATRL w/o BR).
In Fig.~\ref{fig:grid_learning_curve}, we can observe that in all the tasks, without the best response to the fixed point, the learning curves of MATRL w/o BR have higher variance and the lowest final scores.
This establishes the importance of the best response to stabilize and improve agents' performance and empirically shows that MATRL has better convergence ability than do the other baselines.
Additionally, without the SIP to select a fixed point, MATRL recovers to ILs with policy prediction (IL-LA)~\cite{zhang2010multi,lola}.
Similarly, the curves of IL-LA have lower final scores, and the convergence speed is not as good as that of MATRL, which suggests that the SIP provides benefits.
MATRL w/o BR has lower variance compared to IL-LA, which reveals that the SIP can stabilize the learning via weak stable fixed point constraints.
Finally, when compared to IL and IL-LA, as shown in Fig.~\ref{fig:run_time}, in two- to four-agent games with 20,000 environment steps and 50 gradient steps, the training time of MATRL is empirically approximately 1.1-1.2 times slower.
Given the significant performance improvement, 
we believe such extra computational cost from the SIP and the best response are acceptable.  
\vspace{-8pt}
\section{Conclusions}
\vspace{-8pt}
We proposed and analyzed the trust region method for multi-agent learning problems, which considers the IID and SIP to meet multi-agent learning objectives.
In practice, based on independent trust payoff learners, we provide a convenient way to approximate a further restricted step size within the SIP via a meta-game.
This approach ensures that MATRL is generalized, flexible, and easily implemented to deal with multi-agent learning problems in general. Our experimental results justify the fact that the MATRL method significantly outperforms ILs using the same configurations and other strong MARL baselines in both continuous and discrete games with varying numbers of agents.

\bibliography{references}

\begin{thebibliography}{10}

\bibitem{shoham2008multiagent}
Yoav Shoham and Kevin Leyton-Brown.
\newblock {\em Multiagent systems: Algorithmic, game-theoretic, and logical
  foundations}.
\newblock Cambridge University Press, 2008.

\bibitem{yang2020overview}
Yaodong Yang and Jun Wang.
\newblock An overview of multi-agent reinforcement learning from game
  theoretical perspective.
\newblock {\em arXiv preprint arXiv:2011.00583}, 2020.

\bibitem{cao2012overview}
Yongcan Cao, Wenwu Yu, Wei Ren, and Guanrong Chen.
\newblock An overview of recent progress in the study of distributed
  multi-agent coordination.
\newblock {\em IEEE Transactions on Industrial informatics}, 9(1):427--438,
  2012.

\bibitem{zhou2020smarts}
Ming Zhou, Jun Luo, Julian Villela, Yaodong Yang, David Rusu, Jiayu Miao,
  Weinan Zhang, Montgomery Alban, Iman Fadakar, Zheng Chen, et~al.
\newblock Smarts: Scalable multi-agent reinforcement learning training school
  for autonomous driving.
\newblock {\em arXiv preprint arXiv:2010.09776}, 2020.

\bibitem{vinyals2019grandmaster}
Oriol Vinyals, Igor Babuschkin, Wojciech~M Czarnecki, Micha{\"e}l Mathieu,
  Andrew Dudzik, Junyoung Chung, David~H Choi, Richard Powell, Timo Ewalds,
  Petko Georgiev, et~al.
\newblock Grandmaster level in starcraft ii using multi-agent reinforcement
  learning.
\newblock {\em Nature}, 575(7782):350--354, 2019.

\bibitem{peng2017multiagent}
Peng Peng, Ying Wen, Yaodong Yang, Quan Yuan, Zhenkun Tang, Haitao Long, and
  Jun Wang.
\newblock Multiagent bidirectionally-coordinated nets: Emergence of human-level
  coordination in learning to play starcraft combat games.
\newblock {\em arXiv preprint arXiv:1703.10069}, 2017.

\bibitem{berner2019dota}
Christopher Berner, Greg Brockman, Brooke Chan, Vicki Cheung, Przemyslaw
  Debiak, Christy Dennison, David Farhi, Quirin Fischer, Shariq Hashme, Chris
  Hesse, et~al.
\newblock Dota 2 with large scale deep reinforcement learning.
\newblock {\em arXiv preprint arXiv:1912.06680}, 2019.

\bibitem{chatterjee2004nash}
Krishnendu Chatterjee, Rupak Majumdar, and Marcin Jurdzi{\'n}ski.
\newblock On nash equilibria in stochastic games.
\newblock In {\em International Workshop on Computer Science Logic}, pages
  26--40. Springer, 2004.

\bibitem{fudenberg1998theory}
Drew Fudenberg, Fudenberg Drew, David~K Levine, and David~K Levine.
\newblock {\em The theory of learning in games}, volume~2.
\newblock MIT press, 1998.

\bibitem{tan1993multi}
Ming Tan.
\newblock Multi-agent reinforcement learning: Independent vs. cooperative
  agents.
\newblock In {\em Proceedings of the tenth international conference on machine
  learning}, pages 330--337, 1993.

\bibitem{Busoniu2010multiagentRL}
Lucian Bu{\c{s}}oniu, Robert Babu{\v{s}}ka, and Bart De~Schutter.
\newblock Multi-agent reinforcement learning: An overview.
\newblock In {\em Innovations in multi-agent systems and applications-1}, pages
  183--221. Springer, 2010.

\bibitem{hernandez2017survey}
Pablo Hernandez-Leal, Michael Kaisers, Tim Baarslag, and Enrique~Munoz de~Cote.
\newblock A survey of learning in multiagent environments: Dealing with
  non-stationarity.
\newblock {\em arXiv preprint arXiv:1707.09183}, 2017.

\bibitem{ppo}
John Schulman, Filip Wolski, Prafulla Dhariwal, Alec Radford, and Oleg Klimov.
\newblock Proximal policy optimization algorithms.
\newblock {\em CoRR}, abs/1707.06347, 2017.

\bibitem{Kakade2002ApproximatelyOA}
Sham~M. Kakade and John Langford.
\newblock Approximately optimal approximate reinforcement learning.
\newblock In Claude Sammut and Achim~G. Hoffmann, editors, {\em Machine
  Learning, Proceedings of the Nineteenth International Conference {(ICML}
  2002), University of New South Wales, Sydney, Australia, July 8-12, 2002},
  pages 267--274. Morgan Kaufmann, 2002.

\bibitem{nash1950equilibrium}
John~F Nash.
\newblock Equilibrium points in n-person games.
\newblock {\em Proceedings of the national academy of sciences}, 36(1):48--49,
  1950.

\bibitem{bowling2004existence}
Michael Bowling and Manuela Veloso.
\newblock Existence of multiagent equilibria with limited agents.
\newblock {\em Journal of Artificial Intelligence Research}, 22:353--384, 2004.

\bibitem{mazumdar2020gradient}
Eric Mazumdar, Lillian~J Ratliff, and S~Shankar Sastry.
\newblock On gradient-based learning in continuous games.
\newblock {\em SIAM Journal on Mathematics of Data Science}, 2(1):103--131,
  2020.

\bibitem{laurent2011world}
Guillaume~J Laurent, La{\"e}titia Matignon, Le~Fort-Piat, et~al.
\newblock The world of independent learners is not markovian.
\newblock {\em International Journal of Knowledge-based and Intelligent
  Engineering Systems}, 15(1):55--64, 2011.

\bibitem{wellman2006methods}
Michael~P Wellman.
\newblock Methods for empirical game-theoretic analysis.
\newblock In {\em AAAI}, pages 1552--1556, 2006.

\bibitem{lanctot2017unified}
Marc Lanctot, Vin{\'{\i}}cius~Flores Zambaldi, Audrunas Gruslys, Angeliki
  Lazaridou, Karl Tuyls, Julien P{\'{e}}rolat, David Silver, and Thore Graepel.
\newblock A unified game-theoretic approach to multiagent reinforcement
  learning.
\newblock In Isabelle Guyon, Ulrike von Luxburg, Samy Bengio, Hanna~M. Wallach,
  Rob Fergus, S.~V.~N. Vishwanathan, and Roman Garnett, editors, {\em Advances
  in Neural Information Processing Systems 30: Annual Conference on Neural
  Information Processing Systems 2017, December 4-9, 2017, Long Beach, CA,
  {USA}}, pages 4190--4203, 2017.

\bibitem{muller2019generalized}
Paul Muller, Shayegan Omidshafiei, Mark Rowland, Karl Tuyls, Julien
  P{\'{e}}rolat, Siqi Liu, Daniel Hennes, Luke Marris, Marc Lanctot, Edward
  Hughes, Zhe Wang, Guy Lever, Nicolas Heess, Thore Graepel, and R{\'{e}}mi
  Munos.
\newblock A generalized training approach for multiagent learning.
\newblock In {\em 8th International Conference on Learning Representations,
  {ICLR} 2020, Addis Ababa, Ethiopia, April 26-30, 2020}. OpenReview.net, 2020.

\bibitem{omidshafiei2019alpha}
Shayegan Omidshafiei, Christos Papadimitriou, Georgios Piliouras, Karl Tuyls,
  Mark Rowland, Jean-Baptiste Lespiau, Wojciech~M Czarnecki, Marc Lanctot,
  Julien Perolat, and Remi Munos.
\newblock $\alpha$-rank: Multi-agent evaluation by evolution.
\newblock {\em Scientific reports}, 9(1):1--29, 2019.

\bibitem{balduzzi2019open}
David Balduzzi, Marta Garnelo, Yoram Bachrach, Wojciech Czarnecki, Julien
  P{\'{e}}rolat, Max Jaderberg, and Thore Graepel.
\newblock Open-ended learning in symmetric zero-sum games.
\newblock In Kamalika Chaudhuri and Ruslan Salakhutdinov, editors, {\em
  Proceedings of the 36th International Conference on Machine Learning, {ICML}
  2019, 9-15 June 2019, Long Beach, California, {USA}}, volume~97 of {\em
  Proceedings of Machine Learning Research}, pages 434--443. {PMLR}, 2019.

\bibitem{yang2019alpha}
Yaodong Yang, Rasul Tutunov, Phu Sakulwongtana, and Haitham~Bou Ammar.
\newblock $\alpha$$\alpha$-rank: Practically scaling $\alpha$-rank through
  stochastic optimisation.
\newblock In {\em Proceedings of the 19th International Conference on
  Autonomous Agents and MultiAgent Systems}, pages 1575--1583, 2020.

\bibitem{jakob2016communication}
Jakob~N. Foerster, Yannis~M. Assael, Nando de~Freitas, and Shimon Whiteson.
\newblock Learning to communicate with deep multi-agent reinforcement learning.
\newblock In Daniel~D. Lee, Masashi Sugiyama, Ulrike von Luxburg, Isabelle
  Guyon, and Roman Garnett, editors, {\em Advances in Neural Information
  Processing Systems 29: Annual Conference on Neural Information Processing
  Systems 2016, December 5-10, 2016, Barcelona, Spain}, pages 2137--2145, 2016.

\bibitem{maddpg}
Ryan Lowe, Yi~Wu, Aviv Tamar, Jean Harb, Pieter Abbeel, and Igor Mordatch.
\newblock Multi-agent actor-critic for mixed cooperative-competitive
  environments.
\newblock In Isabelle Guyon, Ulrike von Luxburg, Samy Bengio, Hanna~M. Wallach,
  Rob Fergus, S.~V.~N. Vishwanathan, and Roman Garnett, editors, {\em Advances
  in Neural Information Processing Systems 30: Annual Conference on Neural
  Information Processing Systems 2017, December 4-9, 2017, Long Beach, CA,
  {USA}}, pages 6379--6390, 2017.

\bibitem{coma1}
Jakob~N. Foerster, Gregory Farquhar, Triantafyllos Afouras, Nantas Nardelli,
  and Shimon Whiteson.
\newblock Counterfactual multi-agent policy gradients.
\newblock In Sheila~A. McIlraith and Kilian~Q. Weinberger, editors, {\em
  Proceedings of the Thirty-Second {AAAI} Conference on Artificial
  Intelligence, (AAAI-18), the 30th innovative Applications of Artificial
  Intelligence (IAAI-18), and the 8th {AAAI} Symposium on Educational Advances
  in Artificial Intelligence (EAAI-18), New Orleans, Louisiana, USA, February
  2-7, 2018}, pages 2974--2982. {AAAI} Press, 2018.

\bibitem{yang2020multi}
Yaodong Yang, Ying Wen, Jun Wang, Liheng Chen, Kun Shao, David Mguni, and
  Weinan Zhang.
\newblock Multi-agent determinantal q-learning.
\newblock In {\em International Conference on Machine Learning}, pages
  10757--10766. PMLR, 2020.

\bibitem{qmix}
Tabish Rashid, Mikayel Samvelyan, Christian~Schr{\"{o}}der de~Witt, Gregory
  Farquhar, Jakob~N. Foerster, and Shimon Whiteson.
\newblock {QMIX:} monotonic value function factorisation for deep multi-agent
  reinforcement learning.
\newblock In Jennifer~G. Dy and Andreas Krause, editors, {\em Proceedings of
  the 35th International Conference on Machine Learning, {ICML} 2018,
  Stockholmsm{\"{a}}ssan, Stockholm, Sweden, July 10-15, 2018}, volume~80 of
  {\em Proceedings of Machine Learning Research}, pages 4292--4301. {PMLR},
  2018.

\bibitem{vdn}
Peter Sunehag, Guy Lever, Audrunas Gruslys, Wojciech~Marian Czarnecki, Vinicius
  Zambaldi, Max Jaderberg, Marc Lanctot, Nicolas Sonnerat, Joel~Z Leibo, Karl
  Tuyls, et~al.
\newblock Value-decomposition networks for cooperative multi-agent learning
  based on team reward.
\newblock In {\em Proceedings of the 17th international conference on
  autonomous agents and multiagent systems}, pages 2085--2087. International
  Foundation for Autonomous Agents and Multiagent Systems, 2018.

\bibitem{witt2020deep}
Christian~Schroeder de~Witt, Bei Peng, Pierre-Alexandre Kamienny, Philip Torr,
  Wendelin Böhmer, and Shimon Whiteson.
\newblock Deep multi-agent reinforcement learning for decentralized continuous
  cooperative control, 2020.

\bibitem{terry2020arcade}
Justin~K Terry and Benjamin Black.
\newblock Multiplayer support for the arcade learning environment.
\newblock {\em arXiv preprint arXiv:2009.09341}, 2020.

\bibitem{shapley1953stochastic}
Lloyd~S Shapley.
\newblock Stochastic games.
\newblock {\em Proceedings of the national academy of sciences},
  39(10):1095--1100, 1953.

\bibitem{littman1994markov}
Michael~L Littman.
\newblock Markov games as a framework for multi-agent reinforcement learning.
\newblock In {\em Machine learning proceedings 1994}, pages 157--163. Elsevier,
  1994.

\bibitem{trpo}
John Schulman, Sergey Levine, Pieter Abbeel, Michael~I. Jordan, and Philipp
  Moritz.
\newblock Trust region policy optimization.
\newblock In Francis~R. Bach and David~M. Blei, editors, {\em Proceedings of
  the 32nd International Conference on Machine Learning, {ICML} 2015, Lille,
  France, 6-11 July 2015}, volume~37 of {\em {JMLR} Workshop and Conference
  Proceedings}, pages 1889--1897. JMLR.org, 2015.

\bibitem{balduzzi2018mechanics}
David Balduzzi, S{\'{e}}bastien Racani{\`{e}}re, James Martens, Jakob~N.
  Foerster, Karl Tuyls, and Thore Graepel.
\newblock The mechanics of n-player differentiable games.
\newblock In Jennifer~G. Dy and Andreas Krause, editors, {\em Proceedings of
  the 35th International Conference on Machine Learning, {ICML} 2018,
  Stockholmsm{\"{a}}ssan, Stockholm, Sweden, July 10-15, 2018}, volume~80 of
  {\em Proceedings of Machine Learning Research}, pages 363--372. {PMLR}, 2018.

\bibitem{letcher2020impossibility}
Alistair Letcher.
\newblock On the impossibility of global convergence in multi-loss
  optimization.
\newblock {\em arXiv preprint arXiv:2005.12649}, 2020.

\bibitem{tuyls2018generalised}
Karl Tuyls, Julien Perolat, Marc Lanctot, Joel~Z Leibo, and Thore Graepel.
\newblock A generalised method for empirical game theoretic analysis.
\newblock In {\em Proceedings of the 17th International Conference on
  Autonomous Agents and MultiAgent Systems}, pages 77--85. International
  Foundation for Autonomous Agents and Multiagent Systems, 2018.

\bibitem{takeshita2012necessity}
Jun-ichi Takeshita and Hidefumi Kawasaki.
\newblock Necessity and sufficiency for the existence of a pure-strategy nash
  equilibrium.
\newblock {\em Journal of the Operations Research Society of Japan},
  55(3):192--198, 2012.

\bibitem{hansen2003reducing}
Nikolaus Hansen, Sibylle~D M{\"u}ller, and Petros Koumoutsakos.
\newblock Reducing the time complexity of the derandomized evolution strategy
  with covariance matrix adaptation (cma-es).
\newblock {\em Evolutionary computation}, 11(1):1--18, 2003.

\bibitem{tuyls2020bounds}
Karl Tuyls, Julien Perolat, Marc Lanctot, Edward Hughes, Richard Everett,
  Joel~Z Leibo, Csaba Szepesv{\'a}ri, and Thore Graepel.
\newblock Bounds and dynamics for empirical game theoretic analysis.
\newblock {\em Autonomous Agents and Multi-Agent Systems}, 34(1):7, 2020.

\bibitem{mertikopoulos2018optimistic}
Panayotis Mertikopoulos, Bruno Lecouat, Houssam Zenati, Chuan{-}Sheng Foo,
  Vijay Chandrasekhar, and Georgios Piliouras.
\newblock Optimistic mirror descent in saddle-point problems: Going the extra
  (gradient) mile.
\newblock In {\em 7th International Conference on Learning Representations,
  {ICLR} 2019, New Orleans, LA, USA, May 6-9, 2019}. OpenReview.net, 2019.

\bibitem{zhang2010multi}
Chongjie Zhang and Victor~R. Lesser.
\newblock Multi-agent learning with policy prediction.
\newblock In Maria Fox and David Poole, editors, {\em Proceedings of the
  Twenty-Fourth {AAAI} Conference on Artificial Intelligence, {AAAI} 2010,
  Atlanta, Georgia, USA, July 11-15, 2010}. {AAAI} Press, 2010.

\bibitem{lola}
Jakob Foerster, Richard~Y Chen, Maruan Al-Shedivat, Shimon Whiteson, Pieter
  Abbeel, and Igor Mordatch.
\newblock Learning with opponent-learning awareness.
\newblock In {\em Proceedings of the 17th International Conference on
  Autonomous Agents and MultiAgent Systems}, pages 122--130. International
  Foundation for Autonomous Agents and Multiagent Systems, 2018.

\bibitem{letcher2018stable}
Alistair Letcher, Jakob~N. Foerster, David Balduzzi, Tim Rockt{\"{a}}schel, and
  Shimon Whiteson.
\newblock Stable opponent shaping in differentiable games.
\newblock In {\em 7th International Conference on Learning Representations,
  {ICLR} 2019, New Orleans, LA, USA, May 6-9, 2019}. OpenReview.net, 2019.

\bibitem{bowling2002multiagent}
Michael Bowling and Manuela Veloso.
\newblock Multiagent learning using a variable learning rate.
\newblock {\em Artificial Intelligence}, 136(2):215--250, 2002.

\bibitem{heusel2017gans}
Martin Heusel, Hubert Ramsauer, Thomas Unterthiner, Bernhard Nessler, and Sepp
  Hochreiter.
\newblock Gans trained by a two time-scale update rule converge to a local nash
  equilibrium.
\newblock In Isabelle Guyon, Ulrike von Luxburg, Samy Bengio, Hanna~M. Wallach,
  Rob Fergus, S.~V.~N. Vishwanathan, and Roman Garnett, editors, {\em Advances
  in Neural Information Processing Systems 30: Annual Conference on Neural
  Information Processing Systems 2017, December 4-9, 2017, Long Beach, CA,
  {USA}}, pages 6626--6637, 2017.

\bibitem{daskalakis2020ind}
Constantinos Daskalakis, Dylan~J Foster, and Noah Golowich.
\newblock Independent policy gradient methods for competitive reinforcement
  learning.
\newblock In H.~Larochelle, M.~Ranzato, R.~Hadsell, M.~F. Balcan, and H.~Lin,
  editors, {\em Advances in Neural Information Processing Systems}, volume~33,
  pages 5527--5540. Curran Associates, Inc., 2020.

\bibitem{zhang2019conv}
Kaiqing Zhang, Zhuoran Yang, and Tamer Basar.
\newblock Policy optimization provably converges to nash equilibria in zero-sum
  linear quadratic games.
\newblock In H.~Wallach, H.~Larochelle, A.~Beygelzimer, F.~dAlch\'{e} Buc,
  E.~Fox, and R.~Garnett, editors, {\em Advances in Neural Information
  Processing Systems}, volume~32. Curran Associates, Inc., 2019.

\bibitem{chen2006computing}
Xi~Chen, Xiaotie Deng, and Shang-Hua Teng.
\newblock Computing nash equilibria: Approximation and smoothed complexity.
\newblock In {\em 2006 47th Annual IEEE Symposium on Foundations of Computer
  Science (FOCS'06)}, pages 603--612. IEEE, 2006.

\bibitem{daskalakis2009complexity}
Constantinos Daskalakis, Paul~W Goldberg, and Christos~H Papadimitriou.
\newblock The complexity of computing a nash equilibrium.
\newblock {\em SIAM Journal on Computing}, 39(1):195--259, 2009.

\bibitem{fabrikant2004complexity}
Alex Fabrikant, Christos Papadimitriou, and Kunal Talwar.
\newblock The complexity of pure nash equilibria.
\newblock In {\em Proceedings of the thirty-sixth annual ACM symposium on
  Theory of computing}, pages 604--612, 2004.

\bibitem{lipton2003playing}
Richard~J Lipton, Evangelos Markakis, and Aranyak Mehta.
\newblock Playing large games using simple strategies.
\newblock In {\em Proceedings of the 4th ACM conference on Electronic
  commerce}, pages 36--41, 2003.

\bibitem{yang2018mean}
Yaodong Yang, Rui Luo, Minne Li, Ming Zhou, Weinan Zhang, and Jun Wang.
\newblock Mean field multi-agent reinforcement learning.
\newblock In Jennifer~G. Dy and Andreas Krause, editors, {\em Proceedings of
  the 35th International Conference on Machine Learning, {ICML} 2018,
  Stockholmsm{\"{a}}ssan, Stockholm, Sweden, July 10-15, 2018}, volume~80 of
  {\em Proceedings of Machine Learning Research}, pages 5567--5576. {PMLR},
  2018.

\bibitem{littman2002efficient}
Michael~L. Littman, Michael~J. Kearns, and Satinder~P. Singh.
\newblock An efficient, exact algorithm for solving tree-structured graphical
  games.
\newblock In Thomas~G. Dietterich, Suzanna Becker, and Zoubin Ghahramani,
  editors, {\em Advances in Neural Information Processing Systems 14 [Neural
  Information Processing Systems: Natural and Synthetic, {NIPS} 2001, December
  3-8, 2001, Vancouver, British Columbia, Canada]}, pages 817--823. {MIT}
  Press, 2001.

\bibitem{antipin2003extragradient}
Anatoly Antipin.
\newblock Extragradient approach to the solution of two person non-zero sum
  games.
\newblock In {\em Optimization and Optimal Control}, pages 1--28. World
  Scientific, 2003.

\bibitem{spar}
Jie Tang, Keiran Paster, , and Pieter Abbeel.
\newblock Equilibrium finding via asymmetric self-play reinforcement learning.
\newblock {\em Deep Reinforcement Learning Workshop NeurIPS 2018}, 2018.

\bibitem{lockhart2019exploitabilitydescent}
Edward Lockhart, Marc Lanctot, Julien P{\'{e}}rolat, Jean{-}Baptiste Lespiau,
  Dustin Morrill, Finbarr Timbers, and Karl Tuyls.
\newblock Computing approximate equilibria in sequential adversarial games by
  exploitability descent.
\newblock In Sarit Kraus, editor, {\em Proceedings of the Twenty-Eighth
  International Joint Conference on Artificial Intelligence, {IJCAI} 2019,
  Macao, China, August 10-16, 2019}, pages 464--470. ijcai.org, 2019.

\bibitem{leslie2005individual}
David~S Leslie and Edmund~J Collins.
\newblock Individual q-learning in normal form games.
\newblock {\em SIAM Journal on Control and Optimization}, 44(2):495--514, 2005.

\bibitem{leslie2003convergent}
David~S Leslie, EJ~Collins, et~al.
\newblock Convergent multiple-timescales reinforcement learning algorithms in
  normal form games.
\newblock {\em The Annals of Applied Probability}, 13(4):1231--1251, 2003.

\bibitem{mazumdar2019finding}
Eric~V Mazumdar, Michael~I Jordan, and S~Shankar Sastry.
\newblock On finding local nash equilibria (and only local nash equilibria) in
  zero-sum games.
\newblock {\em arXiv preprint arXiv:1901.00838}, 2019.

\bibitem{wen2018probabilistic}
Ying Wen, Yaodong Yang, Rui Luo, Jun Wang, and Wei Pan.
\newblock Probabilistic recursive reasoning for multi-agent reinforcement
  learning.
\newblock In {\em International Conference on Learning Representations}, 2018.

\bibitem{wen2019modelling}
Ying Wen, Yaodong Yang, Rui Luo, and Jun Wang.
\newblock Modelling bounded rationality in multi-agent interactions by
  generalized recursive reasoning.
\newblock {\em arXiv preprint arXiv:1901.09216}, 2019.

\bibitem{lin2020finite}
Tianyi Lin, Zhengyuan Zhou, Panayotis Mertikopoulos, and Michael~I. Jordan.
\newblock Finite-time last-iterate convergence for multi-agent learning in
  games.
\newblock In {\em Proceedings of the 37th International Conference on Machine
  Learning, {ICML} 2020, 13-18 July 2020, Virtual Event}, volume 119 of {\em
  Proceedings of Machine Learning Research}, pages 6161--6171. {PMLR}, 2020.

\bibitem{jordan2009generalization}
Patrick~R Jordan and Michael~P Wellman.
\newblock Generalization risk minimization in empirical game models.
\newblock In {\em Proceedings of The 8th International Conference on Autonomous
  Agents and Multiagent Systems-Volume 1}, pages 553--560, 2009.

\bibitem{dqn}
Volodymyr Mnih, Koray Kavukcuoglu, David Silver, Alex Graves, Ioannis
  Antonoglou, Daan Wierstra, and Martin Riedmiller.
\newblock Playing atari with deep reinforcement learning.
\newblock {\em arXiv preprint arXiv:1312.5602}, 2013.

\bibitem{ddpg}
Timothy~P. Lillicrap, Jonathan~J. Hunt, Alexander Pritzel, Nicolas Heess, Tom
  Erez, Yuval Tassa, David Silver, and Daan Wierstra.
\newblock Continuous control with deep reinforcement learning.
\newblock In Yoshua Bengio and Yann LeCun, editors, {\em 4th International
  Conference on Learning Representations, {ICLR} 2016, San Juan, Puerto Rico,
  May 2-4, 2016, Conference Track Proceedings}, 2016.

\bibitem{sukhbaatar2016learning}
Sainbayar Sukhbaatar, Arthur Szlam, and Rob Fergus.
\newblock Learning multiagent communication with backpropagation.
\newblock In Daniel~D. Lee, Masashi Sugiyama, Ulrike von Luxburg, Isabelle
  Guyon, and Roman Garnett, editors, {\em Advances in Neural Information
  Processing Systems 29: Annual Conference on Neural Information Processing
  Systems 2016, December 5-10, 2016, Barcelona, Spain}, pages 2244--2252, 2016.

\bibitem{bicnet}
Peng Peng, Ying Wen, Yaodong Yang, Quan Yuan, Zhenkun Tang, Haitao Long, and
  Jun Wang.
\newblock Multiagent bidirectionally-coordinated nets: Emergence of human-level
  coordination in learning to play starcraft combat games.
\newblock {\em arXiv preprint arXiv:1703.10069}, 2017.

\bibitem{li2021dealing}
Wenhao Li, Xiangfeng Wang, Bo~Jin, Junjie Sheng, and Hongyuan Zha.
\newblock Dealing with non-stationarity in multi-agent reinforcement learning
  via trust region decomposition.
\newblock {\em arXiv preprint arXiv:2102.10616}, 2021.

\bibitem{li2020multi}
Hepeng Li and Haibo He.
\newblock Multi-agent trust region policy optimization.
\newblock {\em arXiv preprint arXiv:2010.07916}, 2020.

\bibitem{qtran}
Kyunghwan Son, Daewoo Kim, Wan~Ju Kang, David Hostallero, and Yung Yi.
\newblock {QTRAN:} learning to factorize with transformation for cooperative
  multi-agent reinforcement learning.
\newblock In Kamalika Chaudhuri and Ruslan Salakhutdinov, editors, {\em
  Proceedings of the 36th International Conference on Machine Learning, {ICML}
  2019, 9-15 June 2019, Long Beach, California, {USA}}, volume~97 of {\em
  Proceedings of Machine Learning Research}, pages 5887--5896. {PMLR}, 2019.

\bibitem{pangallo2017taxonomy}
Marco Pangallo, James Sanders, Tobias Galla, and Doyne Farmer.
\newblock A taxonomy of learning dynamics in 2 x 2 games, 2017.

\bibitem{singh2000nash}
Satinder Singh, Michael Kearns, and Yishay Mansour.
\newblock Nash convergence of gradient dynamics in general-sum games.
\newblock In {\em Proceedings of the Sixteenth conference on Uncertainty in
  artificial intelligence}, pages 541--548. Morgan Kaufmann Publishers Inc.,
  2000.

\bibitem{magym}
Anurag Koul.
\newblock {A collection of multi agent environments based on OpenAI gym}, 2019.

\bibitem{neng}
Petr Šebek.
\newblock Nash equilibria noncooperative games., sept 2013.

\end{thebibliography}
\bibliographystyle{unsrt}

\onecolumn
\appendix

\setcounter{page}{1}

\addcontentsline{toc}{section}{Appendix} 
\part{{\Large{Appendix for \emph{"A Game-Theoretic Approach to Multi-Agent Trust Region Optimization"}}}} 
\parttoc

\counterwithin{theorem}{section}
\counterwithin{definition}{section}
\counterwithin{lemma}{section}
\counterwithin{remark}{section}

\section{MATRL Algorithm Based on PPO}
\label{app:matrl_ppo}
\begin{algorithm}[ht!]
\begin{algorithmic}[1]
\INPUT The initial policy parameters $\theta_1, \theta_2$, initial value function parameters $\phi_1, \phi_2$ and $\epsilon$.
    \FOR{$k \in \{0,1,2,\cdots\}$} 
      \STATE Using $\pi_1(\theta_1)$, $\pi_2(\theta_2)$ to collect trajectories $\bs{\tau}_1, \bs{\tau}_2$. 
      \STATE Compute GAE reward $\hat{R}_i$ for each $i$.
      \STATE Compute estimated advantages $\hat{A}_1, \hat{A}_2$ based on the current value functions $V_{\phi_1},V_{\phi_2}$.
      \FOR{$i \in \{1,2\}$}
        \STATE Compute a trust payoff region policy $\hat{\pi}_i$ using Eq.~\ref{eq:pi_hat}. 
        \STATE Update the policy by maximizing the PPO-Clip objective:\\
        \resizebox{0.91\hsize}{!}{
        $
        \hat{\theta}_i=\arg \max _{\theta_i} \frac{1}{\left|\bs{\tau}_i\right| T} \sum_{\tau \in \bs{\tau}_i} \sum_{t=0}^{T} \min \left(\frac{\pi_i\left(a_{t} | s_{t};{\theta}\right)}{\pi_i\left(a_{1,t} | s_{t};\theta_i\right)} A_i^{\pi_1,\pi_2}\left(s_{t}, a_{1,t}, a_{2,t}\right), \quad g\left(\epsilon, A_i^{\pi_1,\pi_2}\left(s_{t}, a_{1,t}, a_{2,t}\right)\right)\right),
        $}\\
        where $g$ is a clipping function.
        \STATE Fit value function by regression on mean-squared error:
        $$
        \phi^{\prime}_i =\arg \min _{\phi_i} \frac{1}{\left|\bs{\tau}_i\right| T} \sum_{\tau \in \bs{\tau}_i} \sum_{t=0}^{T}\left(V_{\phi}\left(s_{t}\right)-\hat{R}_{i,t}\right)^{2}
        $$
      \ENDFOR
        \STATE Construct the meta-game $\mathcal{M}(\pi_1(\theta_1), \hat{\pi}_1(\hat{\theta}_1), \pi_2(\theta_2), \hat{\pi}_2(\hat{\theta}_2))$.
        \STATE Solve $\mathcal{M}$ and obtain meta Nash $\rho_1,\rho_2$.
    \STATE Compute aggregated weak stable fixed point $(\bar{\pi}_1$, $\bar{\pi}_2)$. 
    \FOR{$i \in \{1,2\}$}
      \STATE Compute $\pi_i^{(\prime)}$ which best responses to $\bar{\pi}_{-i}$ using Eq.~\ref{eq:br}.
      \STATE Estimate the best response by importance sampling:
        $$
        \theta_i^{\prime}=\frac{\hat{\theta}_i}{\left|\bs{\tau}_i\right| T} \sum_{\tau \in \bs{\tau}_i} \sum_{t=0}^{T} g\left(\epsilon, \pi_i / \bar{\pi}_{-i}\right)
        $$
    \ENDFOR
    \STATE $\theta_1 \leftarrow \theta_1^{\prime}$, $\theta_2 \leftarrow \theta_2^{\prime}$ .   
    \ENDFOR
\OUTPUT $\pi_1(\theta_1)$, $\pi_2(\theta_2)$.
\end{algorithmic}
\caption{Multi-Agent Trust Region Learning Algorithm (PPO Based, Two-Agent Example).}
\label{algo:detial_ppo_matrl}
\end{algorithm}

\section{Independent Trust Payoff Region}
\label{kl_trust_region_proof}

%
%
%
%
%
%
%
%
We use the total variation divergence,
which is defined by $D_{\mathrm{TV}}(p \| q)=\frac{1}{2} \sum_{j}\left|p_{j}-q_{j}\right|$ for discrete probability distributions $p,q$~\cite{trpo}. $D_{\mathrm{TV}}^{\max }(\pi, \tilde{\pi})$ is defined as:
\begin{equation}
D_{\mathrm{TV}}^{\max }(\pi, \tilde{\pi})=\max _{s} D_{\mathrm{TV}}(\pi(\cdot | s) \| \tilde{\pi}(\cdot | s)).
\end{equation} 
Based on this, we can define $\alpha$-coupled policy as:
\begin{definition}[$\alpha$-Coupled Policy~\cite{trpo}]
$(\pi, \pi^{\prime})$ is an $\alpha$-coupled policy pair if it defines a joint distribution $(a,a^{\prime})|s$, such that $P(a \neq a^{\prime} |s) \leq \alpha$ for all $s$. $\pi$ and $\pi^{\prime}$ will denote the marginal distributions of $a$ and $a^{\prime}$, respectively. 
\end{definition}
%
%
%
%
%
%
%
%
When the joint policy pair $\pi_{i}, \pi_{-i}$ changes to $\pi^{\prime}_{i}, \pi^{\prime}_{-i}$ and coupled with $\alpha_{i}$ and $\alpha_{-i}$ correspondingly: 
\begin{equation}
\eta_{i}(\pi_{i}^{\prime}, \pi_{-i}^{\prime})-\eta_{i}(\pi_{i}, \pi_{-i}) \geq 
A_{i}^{\pi_{i}, \pi_{-i}}(\pi_{i}^{\prime}, \pi^{\prime}_{-i})-\frac{4 \gamma \epsilon}{(1-\gamma)^2}(\alpha_{i}+\alpha_{-i}-\alpha_{i}\alpha_{-i})^2,
\end{equation}
where 
$$\epsilon =\max_{s, a_{i}, a_{-i}} \big| A_{i}^{\pi_{i}, \pi_{-i}}(s, a_{i}, a_{-i})\big|.$$
The proofs are as following:
\begin{lemma}[]
     Given that $(\pi_{i}, \pi_{i}^{\prime})$ and $(\pi_{-i}, \pi_{-i}^{\prime})$ are both $\alpha$-coupled policies bounded by $\alpha_{i}$ and $\alpha_{-i}$ respectively, for all $s$,
    \begin{equation}
    \left| A_{i}^{\pi_{i}, \pi_{-i}}\left(s\right)\right| \leq 2(\alpha_{i}+\alpha_{-i}-\alpha_{i}\alpha_{-i}) \max_{s, a_{-i}, a_{-i}} \big| A_{i}^{\pi_{i}, \pi_{-i}}(s, a_{i}, a_{-i})\big|
    \end{equation}
\begin{proof}
\begin{align}
A_{i}^{\pi_{i}, \pi_{-i}}(s) &= \mathbb{E}_{a_{i}^{\prime}, a_{-i}^{\prime} \sim \pi_{i}^{\prime}, \pi^{\prime}_{-i}}    \left[A_{i}^{\pi_{i}, \pi_{-i}}(s,a_{i}^{\prime}, a_{-i}^{\prime})\right]\\
&= \mathbb{E}_{(a_{i},a_{i}^{\prime}) \sim (\pi_{i}, \pi_{i}^{\prime}), (a_{-i}, a_{-i}^{\prime}) \sim (\pi_{-i}, \pi^{\prime}_{-i})}   \left[A_{i}^{\pi_{i}, \pi_{-i}}(s,a_{i}^{\prime}, a_{-i}^{\prime})-A_{i}^{\pi_{i}, \pi_{-i}}(s,a_{i}, a_{-i})\right]\\
&=P(a_{i}\neq a_{i}^{\prime} \lor a_{-i}\neq a_{-i}^{\prime}|s)\mathbb{E}_{(a_{i},a_{i}^{\prime}) \sim (\pi_{i}, \pi_{i}^{\prime}), (a_{-i}, a_{-i}^{\prime}) \sim (\pi_{-i}, \pi^{\prime}_{-i})} [A_{i}^{\pi_{i}, \pi_{-i}}(s,a_{i}^{\prime}, a_{-i}^{\prime})\\
&\qquad\qquad\qquad\qquad\qquad\qquad\qquad\qquad\qquad\qquad\qquad\qquad-A_{i}^{\pi_{i}, \pi_{-i}}(s,a_{i}, a_{-i})]\\
&\leq (\alpha_{i}+\alpha_{-i}-\alpha_{i}\alpha_{-i}) \cdot  2  \max_{s, a_{-i}, a_{-i}} \big| A_{i}^{\pi_{i}, \pi_{-i}}(s, a_{i}, a_{-i})\big|,
\end{align}
where $P(a_{i}\neq a_{i}^{\prime} \lor a_{-i}\neq a_{-i}^{\prime}|s)=1-(1-\alpha_{i})(1-\alpha_{-i})=\alpha_{i}+\alpha_{-i}-\alpha_{i}\alpha_{-i}$.

\end{proof}
\end{lemma}

\begin{lemma}
    Let $(\pi_{i}, \pi_{i}^{\prime})$ and $(\pi_{-i}, \pi_{-i}^{\prime})$ are  $\alpha$-coupled policy pairs. Then,
    \begin{equation}
    \begin{aligned}
        &\Big|\mathbb{E}_{s_t\sim\pi_{i}^{\prime},\pi_{-i}^{\prime}}\big[A_{i}^{\pi_{i}, \pi_{-i}}\left(s\right)\big] - \mathbb{E}_{s_t\sim\pi_{i},\pi_{-i}}\big[A_{i}^{\pi_{i}, \pi_{-i}}\left(s\right)\big] \Big| \\
        &\qquad\qquad\leq 4(\alpha_{i}+\alpha_{-i}-\alpha_{i}\alpha_{-i})(1-(1-\alpha_{i})^t(1-\alpha_{-i})^t)\max_{s, a_{-i}, a_{-i}} \big| A_{i}^{\pi_{i}, \pi_{-i}}(s, a_{i}, a_{-i})\big|
    \end{aligned}
    \end{equation}
    \begin{proof}
        The preceding Lemma bounds the difference in expected advantage at each time step $t$. 
When $t^{\prime} = 0$ indicates that $\pi_{i},\pi_{-i}$ and $\pi_{i}^{\prime},\pi_{-i}^{\prime}$ both agreed on all time steps less than $t$.
By the definition of $\alpha_{i},\alpha_{-i}$, $P(\pi_{i},\pi_{-i}:=\pi_{i}^{\prime},\pi_{-i}^{\prime} | t=i) \geq (1-\alpha_{i})(1-\alpha_{-i})$, so $P(t^{\prime}=0) \geq (1-\alpha_{i})^t(1-\alpha_{-i})^t$ and $P(t^{\prime}>0) \leq 1-(1-\alpha_{i})^t(1-\alpha_{-i})^t$.
We can sum over time to bind the
difference between $\eta_{i}(\pi_{i}^{\prime},\pi_{-i}^{\prime})$ and $\eta_{i}(\pi_{i}, \pi_{-i})$.
\begin{align}
\Big|\eta_{i}(\pi_{i}^{\prime},\pi_{-i}^{\prime}) - L_{i}^{\pi_{i}, \pi_{-i}}(\pi_{i}^{\prime}, \pi_{-i}^{\prime}) \Big| 
&= \sum_{t=0}^{\infty}\gamma^t\Big|\mathbb{E}_{s_t\sim\pi_{i}^{\prime},\pi_{-i}^{\prime}}\big[A_{i}^{\pi_{i}, \pi_{-i}}\left(s\right)\big] - \mathbb{E}_{s_t\sim\pi_{i},\pi_{-i}}\big[A_{i}^{\pi_{i}, \pi_{-i}}\left(s\right)\big] \Big|\\
&\leq \sum_{t=0}^{\infty}\gamma^t \cdot 4\epsilon(\alpha_{i}+\alpha_{-i}-\alpha_{i}\alpha_{-i})(1-(1-\alpha_{i})^t(1-\alpha_{-i})^t) \\
&= 4\epsilon(\alpha_{i}+\alpha_{-i}-\alpha_{i}\alpha_{-i})\Big(\frac{1}{1-\gamma} - \frac{1}{1-\gamma(1-\alpha_{i})(1-\alpha_{-i})} \Big)\\
&= \frac{4\epsilon(\alpha_{i}+\alpha_{-i}-\alpha_{i}\alpha_{-i})^2}{(1-\gamma)(1-\gamma(1-\alpha_{i})(1-\alpha_{-i}))}\\
&\leq \frac{4\epsilon(\alpha_{i}+\alpha_{-i}-\alpha_{i}\alpha_{-i})^2}{(1-\gamma)^2},
\end{align}
where $\epsilon =\max_{s, a_{i}, a_{-i}} \big| A_{i}^{\pi_{i}, \pi_{-i}}(s, a_{i}, a_{-i})\big|$.
    \end{proof}
\end{lemma}
Note that
\begin{equation}
L_{i}^{\pi_{i}, \pi_{-i}}(\pi_{i}^{\prime}, \pi_{-i}^{\prime}) = \eta_{i}(\pi_{i}, \pi_{-i}) + \sum_s{\rho^{\pi_{i}, \pi_{-i}}(s)} \sum_{a_{i}}{\pi_{i}^{\prime}(a_{i}|s)} \sum_{a_{-i}}{\pi_{-i}^{\prime}(a_{-i}|s)} A_{i}^{\pi_{i}, \pi_{-i}} (s, a_{i}, a_{-i}).
\end{equation}
Then, we can have
\begin{equation}
\eta_{i}(\pi_{i}^{\prime}, \pi_{-i}^{\prime})-\eta_{i}(\pi_{i}, \pi_{-i}) \geq 
A_{i}^{\pi_{i}, \pi_{-i}}(\pi_{i}^{\prime}, \pi^{\prime}_{-i})-\frac{4 \gamma \epsilon}{(1-\gamma)^2}(\alpha_{i}+\alpha_{-i}-\alpha_{i}\alpha_{-i})^2.
\end{equation}

\section{Proof of Theorem 1}
\label{app:stable_point}
At each iteration, denote $\nabla_{i} g_{i} = \nabla_{\mathbf{\pi}_{i}} g_{i}^{\pi_i,\pi_{-i}}$ and $\nabla_{i,-i} g_{i} = \nabla_{\mathbf{\pi}_{i}}\nabla_{\mathbf{\pi}_{-i}} g_{i}^{\pi_i,\pi_{-i}}$ for each $i$. Consider the simultaneous gradient $\boldsymbol{\xi}$ of the expected advantage gains and the corresponding Hessian $H$:
\begin{equation}
\label{eq:meta_sim_grad}
  \boldsymbol{\xi}(\pi_i, \pi_{-i})=\left(\nabla_{i} g_{i},\nabla_{-i} g_{-i}\right),
\end{equation}
\begin{equation}
\label{eq:meta_hessian}
  H=\nabla \xi=\left(\begin{array}{cc}\nabla_{i,i} g_{i}  & \nabla_{i,-i} g_{i}  \\ \ \nabla_{-i,i} g_{-i} & \nabla_{-i,-i} g_{-i}\end{array}\right).
\end{equation}
For a restricted underlying game, where policy space is bounded: $\pi_i \in [\pi_i,\hat{\pi}_i]$. Assume $\pi_i$ is the linear mixture of $\pi_i,\hat{\pi}_i$, and $\bar{\pi}_i = \rho_i \pi_i + (1-\rho_i) \hat{\pi}_i$, where $\rho_i \in [0,1]$. Therefore, we can re-write the $g_{i}^{\pi_i,\pi_{-i}}(\pi_i,\pi_{-i})$ in the form of:
\begin{equation}
  g_{i}^{\pi_i,\pi_{-i}}(\pi_i,\pi_{-i})= g_{i}^{\pi_i,\pi_{-i}}(\rho_i,\rho_{-i}) = \rho_i(1-\rho_{-i})g_{i}^{i,-\hat{i}}+(1-\rho_i)\rho_{-i}g_{i}^{\hat{i},-i}+(1-\rho_i)(1-\rho_{-i})g_{i}^{\hat{i},-\hat{i}}.
\end{equation}
Then we have:
\begin{equation}
  \nabla_{i} g_{i}(\rho_{-i})=(1-\rho_{-i})g_{i}^{i,-\hat{i}}-\rho_{-i}g_{i}^{\hat{i},-i}-(1-\rho_{-i})g_{i}^{\hat{i},\hat{-i}},
\end{equation}
and $\boldsymbol{\xi}(\pi_i, \pi_{-i})=\boldsymbol{\xi}(\rho_i, \rho_{-i})$.
Given a meta Nash policy pair $(\bar{\pi}_i,\bar{\pi}_{-i})$, where $\bar{\pi}_i = \bar{\rho}^i \pi_i + (1-\bar{\rho}^i ) \hat{\pi}_{i}$, according to the Nash definition, we have:
\begin{equation}
\begin{pmatrix}
\bar{\rho}_i \\
1-  \bar{\rho}_i
\end{pmatrix}^T
\begin{pmatrix}
g_{i}^{i,-i} & g_{i}^{i,-\hat{i}} \\
g_{i}^{\hat{i},-i} & g_{i}^{\hat{i},-\hat{i}}
\end{pmatrix}
\begin{pmatrix}
\bar{\rho}_{-i} \\
1-  \bar{\rho}_{-i}
\end{pmatrix} 
\geq 
\begin{pmatrix}
\rho_i \\
1-  \rho_i
\end{pmatrix}^T
\begin{pmatrix}
g_{i}^{i,-i} & g_{i}^{i,-\hat{i}} \\
g_{i}^{\hat{i},-i} & g_{i}^{\hat{i},-\hat{i}}
\end{pmatrix}
\begin{pmatrix}
\bar{\rho}_{-i} \\
1-  \bar{\rho}_{-i}
\end{pmatrix},
\end{equation}

which implies:


\begin{equation}
\label{eq:meta_nash_cond}
  \begin{aligned}
  (\bar{\rho}_i - \rho_i)\nabla_{i} g_{i}(\bar{\rho}_{-i})&\geq 0, \quad \bar{\rho}_i, \forall \rho_{-i} \in [0,1],   \\
  (\bar{\rho}_{-i} - \rho_{-i})\nabla_{-i} g_{-i}(\bar{\rho}_{i})&\geq 0, \quad \bar{\rho}_{i}, \forall \rho_{-i} \in [0,1].
\end{aligned}
\end{equation}
When $\bar{\rho}_i,\bar{\rho}_{-i} \in (0,1)$ in accordance with the Nash condition in Eq.~\ref{eq:meta_nash_cond}, $\nabla_{i}g_{i}(\bar{\rho}_{-i})=\nabla_{-i}g_{-i}(\bar{\rho}_{i})=0$. It shows that $(\bar{\pi}_i,\bar{\pi}_{-i})$ is a fixed point due to $\boldsymbol{\xi}(\bar{\pi}_i,\bar{\pi}_{-i}) = \boldsymbol{\xi}(\bar{\rho}_i,\bar{\rho}_{-i})  = \bs{0}$. For the boundary case, where $\bar{\rho}_i$ or $\bar{\rho}_{-i} \in \{0,1\}$, because they are constrained to the unit square $[0,1]\times[0,1]$, the gradients on the boundaries of the unit square are projected onto the unit square, which means additional points of zero gradient exist. In other words, $\nabla_{i}g_{i}$ and $\nabla_{-i}g_{-i}$ are still equal to zero in boundary case, and the $(\bar{\pi}_i,\bar{\pi}_{-i})$ is a fixed point in both cases.

Next, we determine what types of the fixed point that $(\bar{\pi}_i,\bar{\pi}_{-i})$ belongs to. 
According to the Eq.~\ref{eq:meta_hessian}, we have the exact Hessian Matrix for the restricted game:
\begin{equation}
H=\nabla \xi=\left(\begin{array}{cc}0  & g_{i}^{\hat{i},-\hat{i}} -g_{i}^{i,-\hat{i}}-g_{i}^{\hat{i},-i}\\ \ g_{-i}^{\hat{i},-\hat{i}} - g_{-i}^{i,-\hat{i}}-g_{-i}^{\hat{i},-i}& 0\end{array}\right)
\end{equation}
The eigenvalue $\lambda$ of $H$ can be computed:
\begin{equation}
\lambda^{2}-\operatorname{Tr}(H) \lambda+\operatorname{det}(H)=\lambda^{2}-(g_{i}^{\hat{i},-\hat{i}} -g_{i}^{i,-\hat{i}}-g_{i}^{\hat{i},-i})(g_{-i}^{\hat{i},-\hat{i}} - g_{-i}^{i,-\hat{i}}-g_{-i}^{\hat{i},-i})=0
\end{equation}
Denotes $\bar{g}_{i} := g_{i}^{\hat{i},-\hat{i}} -g_{i}^{i,-\hat{i}}-g_{i}^{\hat{i},-i}$, we have
$\bs{\lambda} = \pm \sqrt{\bar{g}_{i}\bar{g}_{-i}}$. Therefore, we can have following cases for the fixed point $(\bar{\rho}_i,\bar{\rho}_{-i})$:
\begin{enumerate}
  \item Fully cooperative games: $\bar{g}_{i} \leq 0, \bar{g}_{-i} \leq 0$, then $H(\bar{\rho}_i,\bar{\rho}_{-i}) \preceq 0$, which means $(\bar{\rho}_i,\bar{\rho}_{-i})$ is a stable fixed point as we are maximizing the objective.
  \item Fully competitive games: $\bar{g}_{i}>0, \bar{g}_{-i}<0$ or $\bar{g}_{i}<0, \bar{g}_{-i}>0$,  all $\bs{\lambda}$ have two pure imaginary eigenvalues with zero real part, where $(\bar{\rho}_i,\bar{\rho}_{-i})$ is a saddle point.
  \item General-sum games: they are in-between the cooperative and competitive games, which means $(\bar{\rho}_i,\bar{\rho}_{-i})$ can be either stable fixed point or saddle point.
\end{enumerate}

Because we assume $\hat{\pi}_i$ monotonically improved compared to $\pi_i$, then even in zero-sum case, there is at least one negative value in $\bar{g}_{i}$ and $\bar{g}_{-i}$. Therefore, in all the situations, $(\bar{\rho}_i,\bar{\rho}_{-i})$ is not unstable, and could be a stable point or saddle point. We define them as a weak stable fixed point. 
It also has a tighter lower bound than the independent trust region improvement seen in Remark~\ref{the:nash_region}:

\begin{remark}
\label{the:nash_region}
Let $(\rho_{i}, \rho_{-i})$ be a Nash equilibrium of the policy-space meta-game $\mathcal{M}(\pi_i, \hat{\pi}_{i}, \pi_{-i}, \hat{\pi}_{-i})$, which is used for computing the linear mixture policies $\bar{\pi}_i, \bar{\pi}_{-i}$. For simplicity, define $\bar{\rho}_i = 1 - \rho_i $, then we have the payoff improvement lower bound for $\bar{\pi}_i, \bar{\pi}_{-i}$:
\begin{equation}
\eta_i(\bar{\pi}_i, \bar{\pi}_{-i})-\eta_i(\pi_{i}, \pi_{-i}) \geq 
g_i^{\pi_i, \pi_{-i}}(\bar{\pi}_i, \bar{\pi}_{-i})-\frac{4 \gamma \epsilon_i}{(1-\gamma)^2}(\alpha_i\bar{\rho}_{i}+\alpha_{-i}\bar{\rho}_{-i}-\alpha_{i}\alpha_{-i}\bar{\rho}_i\bar{\rho}_{-i})^2,
\end{equation}
\vspace{-5pt}
that is a tighter lower bound compared with Theorem~\ref{the:payoff_region}.
\end{remark} 

Finally, we obtain MATRL as follows: First, an agent $i$ collects a set of trajectories using its current policy $\pi_i$ by independent play with other agents. Then a predicted policy $\hat{\pi}_i$ can be estimated using the single-agent trust region methods, which has a trust payoff improvement against the other agents' current policy $\pi_{-i}$. 
However, this trust payoff improvements would not benefit convergence requirements for the multi-agent system due to other agents adaptive learning. 
To solve this problem, we approximate a $n$-agent two-action meta-game in policy-space by reusing the trajectories from the last TPR step.
In this game, each agent $i$ has two pure strategies: choosing the \emph{current policy} $\pi_i$ or \emph{predicted policy} $\hat{\pi}_i$ and the corresponding payoffs are the expected advantages (defined in Eq.~\ref{eq:adv})
of the joint policy pairs. By constructing such a meta-game, we transform a complex multi-agent interactions problem into game-theoretic analysis concerning the underlying game restricted in $[\pi_i,\hat{\pi}_i]$. 
Then we can obtain a weak stable fixed point as TSR within the TPR by solving the meta-game,.
When the fixed point is a saddle point we then take the best response to the weak stable fixed point to get the next iteration’s policies.  
This encourages exploration and avoid stagnation at an unexpected saddle point.  

\section{Proof of Theorem 2}
\label{app:the2_proof}

Let the objectives $\eta_{i}(\pi_1, \cdots ,\pi_{n})$ of agents are twice continuously differentiable, in which the agents with parameters $\theta=\left(\theta^{1}, \ldots, \theta^{n}\right)$.
Denote $\xi$ as the simultaneous gradient of game, we can obtain the corresponding Hession of the game:
$$
H=\nabla \xi=\left(\begin{array}{ccc}\nabla_{11} \eta_{1} & \cdots & \nabla_{1 n} \eta_{1} \\ \vdots & \ddots & \vdots \\ \nabla_{n 1} \eta_{n} & \cdots & \nabla_{n n} \eta_{n}\end{array}\right).
$$
Let $H_{o}$ is the matrix of anti-diagonal blocks of $H$ (Hessian of the game), and $\alpha$ is step-size. For MATRL, we have the updating gradient $\left(I-\rho\alpha H_{o}\right) \xi$, where $\rho$ is a ratio determined by meta-game Nash to dynamically adjust the step-size at each iteration. Then the iterative procedure:
$$
F(\theta)=\theta+\alpha (I-\rho\alpha H_{o}) \xi(\theta).
$$
Assume $\bar{\theta}$ is a fixed point if $\xi(\bar{\theta})=0$, and denote $X := (I-\rho\alpha H_{o}) $, then we have:
$$
\nabla[X \xi](\bar{\theta})=\nabla X(\bar{\theta}) \xi(\bar{\theta})+X(\bar{\theta}) \nabla \xi(\bar{\theta})=X H(\bar{\theta})
$$
is negative stable (if all its eigenvalues of $X$ have negative real part) according to Theorem 1,  namely has eigenvalues $a_{k}+i b_{k}$ with $a_{k}<0$. It means 
$$\nabla F(\bar{x})=I+\alpha \nabla[X \xi](\bar{x})$$
has eigenvalues $1+ \alpha a_{k}+i \alpha b_{k}$ in the a small circle with radius $\epsilon \geq 0$:
$$\left|1+\alpha a_{k}+i \alpha b_{k}\right|^{2}<1   \Longleftrightarrow \quad 0<\alpha<\frac{-2 a_{k}}{a_{k}^{2}+b_{k}^{2}} ,$$
which is always possible for $a_{k}<0$. Then it is sufficient to prove the converges locally to $\bar{\theta}$ with $\epsilon$ error for $\alpha$ sufficiently small according to Ostrowski's Theorem~\cite{letcher2018stable}.

\section{Environment Details}
\label{app:exp_env_detail}

\begin{figure}[t!]
\sbox\twosubbox{%
  \resizebox{\dimexpr.99\textwidth-1em}{!}{%
  \includegraphics[height=.5\textwidth]{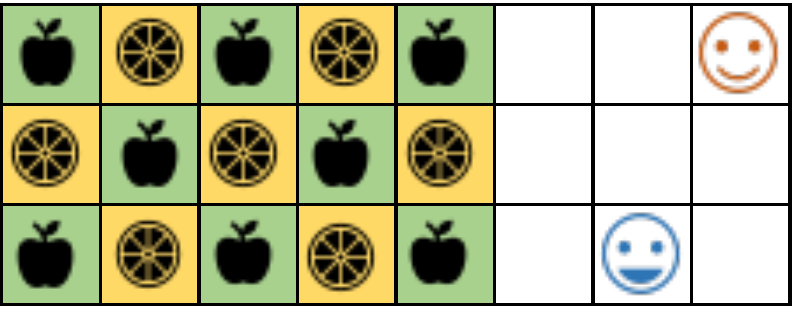}%
  \includegraphics[height=.5\textwidth]{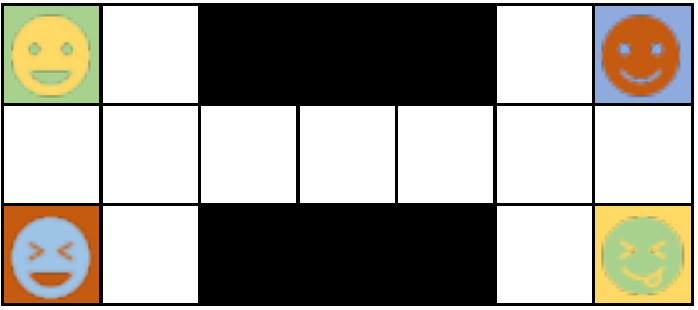}%
    \includegraphics[height=.5\textwidth]{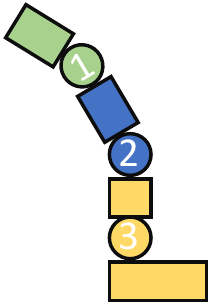}%
  }%
}
\setlength{\twosubht}{\ht\twosubbox}
\centering
\subcaptionbox{}{%
  \label{fig:checker_game}
  \includegraphics[height=\twosubht]{./figures/checker_game}%
}
\subcaptionbox{}{%
  \label{fig:switch4_game}
  \includegraphics[height=\twosubht]{./figures/switch4_game}%
}
\subcaptionbox{}{%
  \label{fig:hopper_game}
  \includegraphics[height=\twosubht]{./figures/hopper_game}%
}
\caption{Multi-agent discrete and continuous action tasks: (a) 2-agent checker (discrete), (b) 4-agent switch (discrete), (c) 3-agent MuJoCo hopper (continious).}
\label{fig:games}
\end{figure}

\begin{figure}[t!]
\vspace{-5pt}
  \centering
  \includegraphics[width=.4\linewidth]{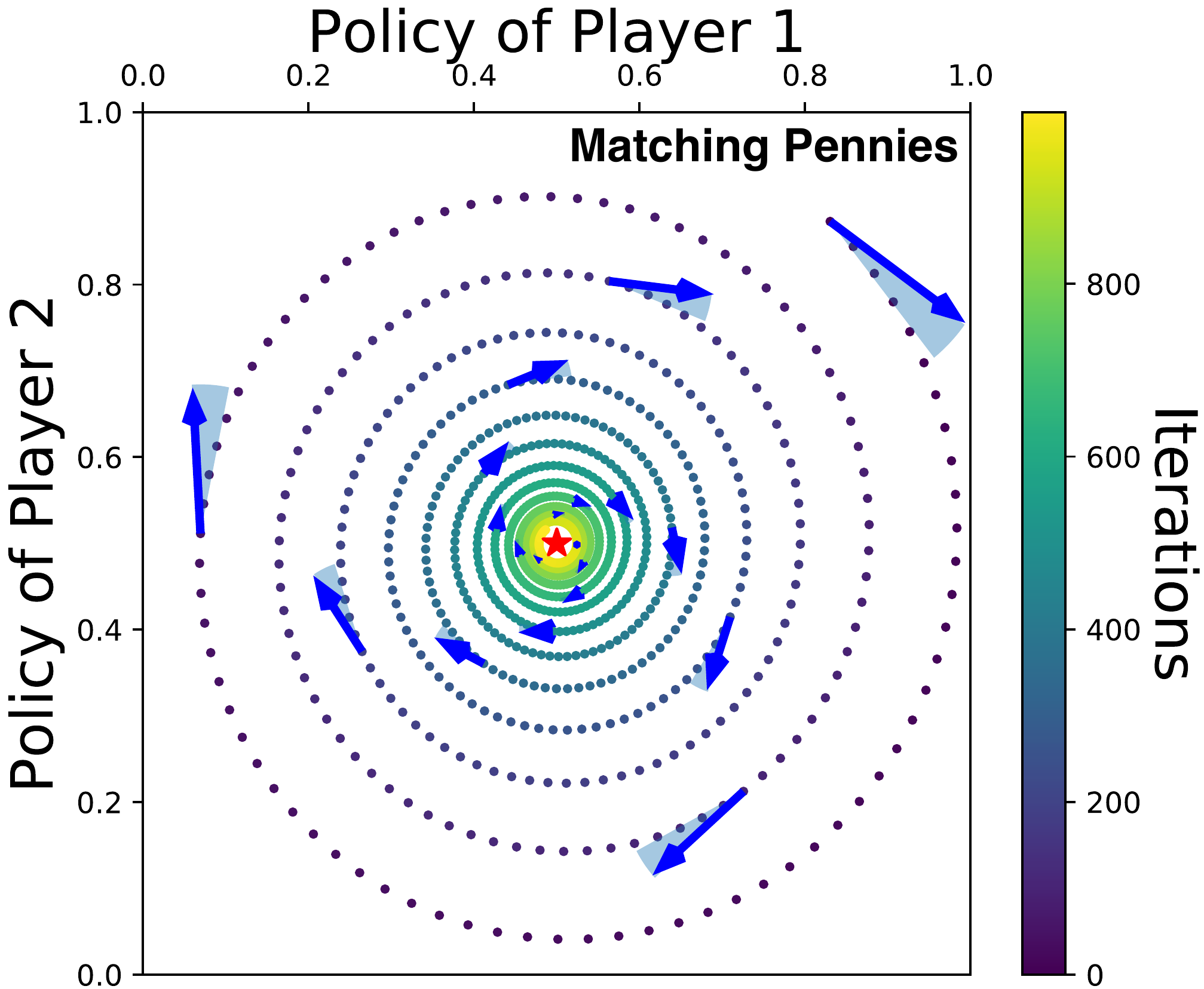}
  \vspace{-5pt}
\caption{Learning the dynamics of MATRL in a matching pennies (MP) game. The blue arrow is the gradient direction, and the pale blue area is the STR. }
    \label{fig:mp}
    \vspace{-15pt}
\end{figure}

\textbf{Random $\mathbf{2 \times 2}$ Matrix Games}.
\label{sec:random_matrix_game}
We created a generator of $2 \times 2$ matrix games based on the category provided by ~\cite{pangallo2017taxonomy}. 
Coordination games have characteristics enabling one agent to improve the payoff without decreasing the payoff of the other agent.
Anti-coordination games are ones where one agent improves the payoff while the other agent’s payoff decreases.
Both coordination and anti-coordination games can have two pure NEs and one mixed strategy NE.  
In cyclic games, the action selections of agents that is based on their actions will form a cycle, ensuring that there is no pure NE in the game.
Instead only mixed strategy NE will be found.

\textbf{Grid World Games}.
In two-player checker, as shown in Fig.~\ref{fig:games}a, there is one sensitive player who gets reward 5 when they collect an apple and 5 when they collect a lemon; a less sensitive player gets 1 for apple and 1 for lemon. 
The learning goal is to let the sensitive player get apples and the other one get lemons to have a higher total reward. 
In four-player switch, as shown in Fig.~\ref{fig:games}b, to reach the targets, agents need to figure out a way to go through a narrow corridor. 
The agent gets $-1$ for taking each step and $5$ when arriving at a target. 
Four-player switch uses the same map as two-player switch, where two agents start from the left side and the others from the right side to go through the corridor to reach the targets. 
With more agents in four-player switch, learning becomes more challenging. MATRL agents achieved higher total rewards compared to baseline algorithms within the same number of steps.

\textbf{Multi-Agent MuJoCo Tasks}.
We used the three-agent Hopper environment described in~\cite{witt2020deep}, and Fig.~\ref{fig:games}c, where three agents control three joints of the robot and learn to cooperate to move forward as far as possible. 
The agent is rewarded by the number of time steps that they move without falling.  
Each agent has 3 continuous output values as the action, and all the agents have a full observation of the states of size $17$. 
We use the same hyper-parameters for MATRL, MATRL w/o BR, and IL-PP. 
For MADDPG agent, we use the hyper-parameters described in the paper~\cite{witt2020deep}.


\textbf{Multi-Agent Atari Game}. The pong game is a multi-agent Atari version\footnote{\url{https://github.com/PettingZoo-Team/Multi-Agent-ALE}} of table tennis
Two players must prevent a ball from whizzing past their paddles and allowing their opponent to score. The game ends when one side earns 21 points.

\begin{figure}[t!]
     \centering
	\begin{subfigure}[r]{.33\textwidth}
         \centering
         \includegraphics[width=\textwidth]{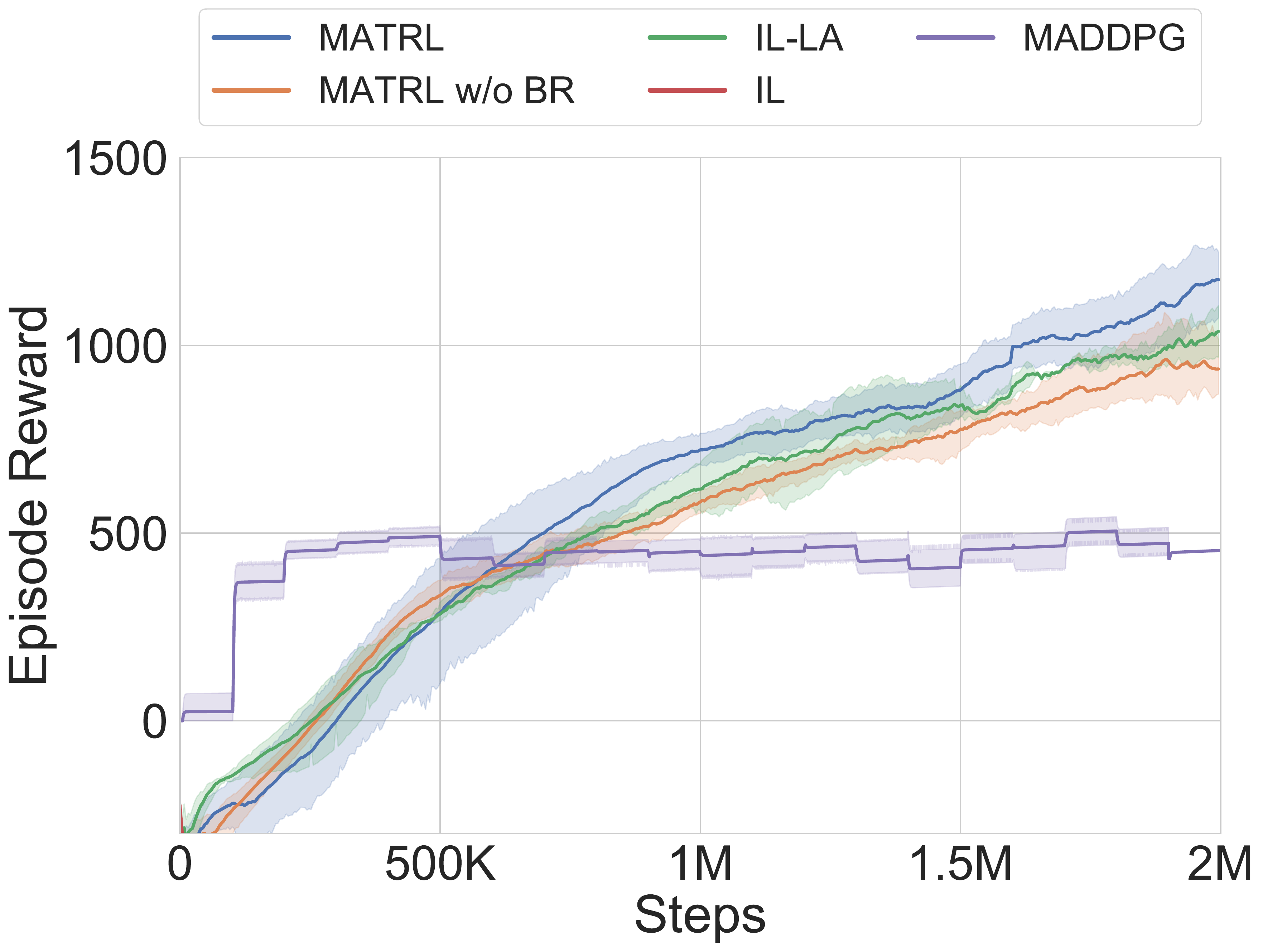}
        \caption{2-Agent Ant.}
        \label{fig:ant}
     \end{subfigure}
     \begin{subfigure}[r]{.33\textwidth}
         \centering
         \includegraphics[width=\textwidth]{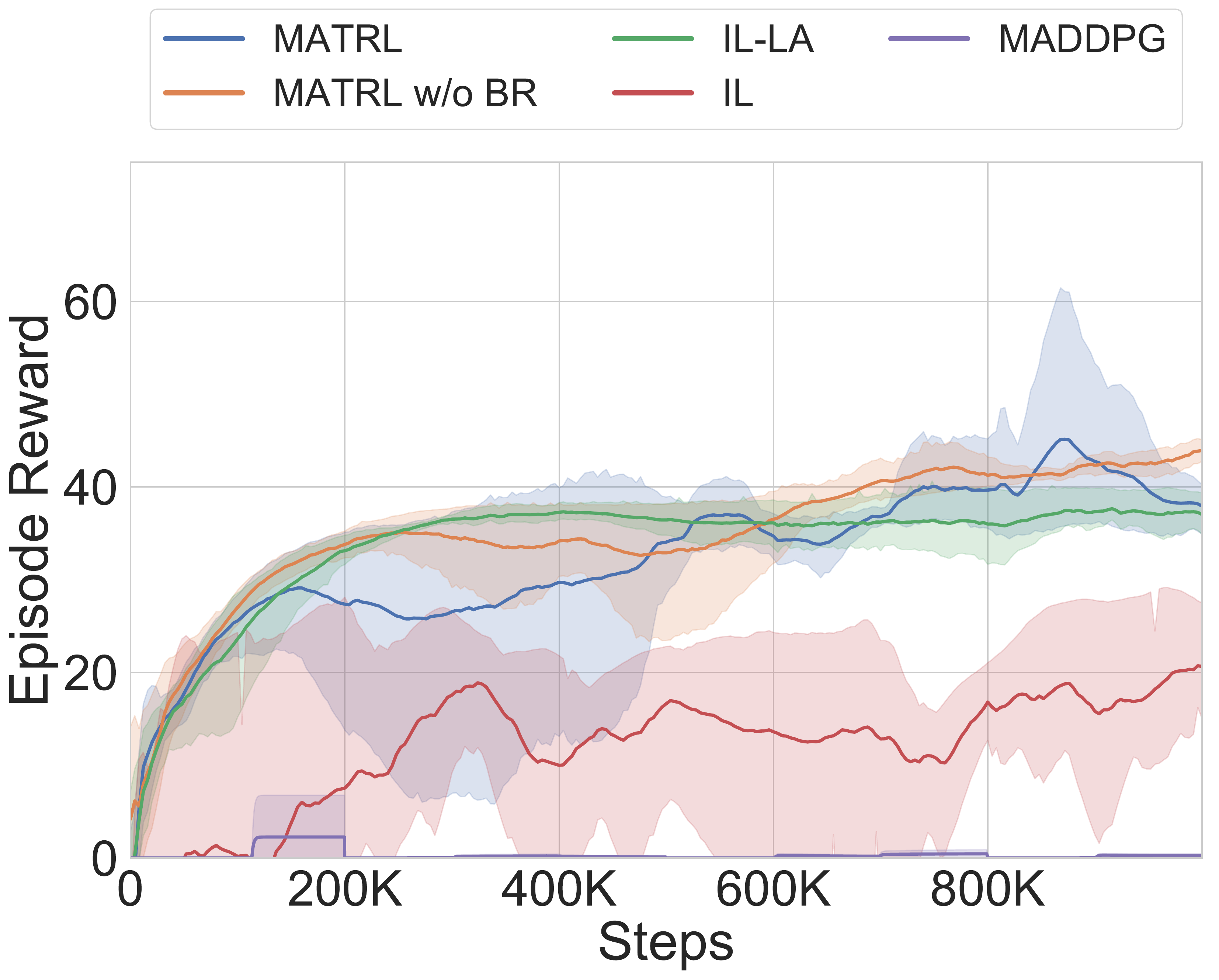}
        \caption{2-Agent Swimmer.}
        \label{fig:Swimmer}
     \end{subfigure}
     \caption{Learning curves of 2-Agent Ant and 2-Agent Swimmer MuJoCo tasks.}
     \label{fig:matrix_extra}
\end{figure}

\section{Experimental Parameter Settings}
\label{app:exp_detail}

 For all the tasks, the most important hyper-parameters are learning rate/step size, the number of update steps, batch size and value, policy loss coefficient. Appropriate learning rate and update steps plus larger batch size give a more stable learning curve. And for different environments, policy and value network loss coefficients that keep two losses at the same scale are essential in improving the learning result and speed. Also, for meta-game construction and best response update where we use the importance ratio to do estimation, a clipping factor of the ration is vital to achieving a stable and monotonic improving result. The followings are the detailed parameter settings for each task.

\textbf{Matrix Game and Random $2 \times 2$ Matrix Games}.
The hyper-parameters settings for MATRL, IGA-PP, and WoLF are listed in Table~\ref{tb:hyper_baselines_matrix}. 
As shown in Fig.~\ref{fig:matrix_extra}, we also listed the additional convergence analysis in classical Chicken and Prisoners' Dilemma Games, which demonstrate good convergence performance of MATRL on both games.
For MATRL, we have the KL-divergence coefficient as an extra hyper-parameter to add the KL-divergence as part of the loss in policy updating. And for the baseline algorithm WoLF, we give the real NE of the game as part of the parameters. In all the games, all the algorithms shared the same initial policy values $[0.9, 0.1]$ for player 1 and $[0.2, 0.8]$ for player 2.

\begin{table}[t!]
\caption{Hyper-parameter settings in $2 \times 2$ matrix games.}
\label{tb:hyper_baselines_matrix}
\centering
\begin{sc}
\resizebox{.9\linewidth}{!}{
\begin{tabular}{lcl}
\toprule
\textbf{Settings }& \textbf{Value} & \textbf{Description}  \\
\midrule 
\multicolumn{3}{l}{\textbf{Common Settings}}  \\  \hline
Initial policies 1 & $[0.9, 0.1]$ & The initial policy values for player 1 \\
Initial policies 2 & $[0.2, 0.8]$ & The initial policy values for player 2 \\ \hline
\multicolumn{3}{l}{\textbf{MATRL Settings}} \\ \hline
Best Response learning rate & 0.03 & The learning rate for the best response step \\
KL coefficient & 100 & The KL-Divergence coefficient in policy loss\\   \hline
\multicolumn{3}{l}{\textbf{WoLF Settings}} \\ \hline
Learning Rate Maximum & 0.06 & The maximum learning rate for WoLF learn fast agent  \\ 
Learning Rate Minimum & 0.02 & The minimum learning rate for WoLF win agent  \\
\bottomrule
\end{tabular}
}
\end{sc}
\end{table}

\textbf{Grid World Games and Multi-agent Continuous Control Task}.
The hyper-parameters settings for MATRL are given in Table~\ref{tb:hyper}.  We used the same hyper-parameters for MATRL, MATRL w/o BR, IL-PP, and IL. The only difference is whether to use Best Response and the meta-game or not.  We used Leaky ReLU as the activation function for both policies and value networks. For the training, we used paralleled workers to collect experience data and update the network weights separated then synchronize all the works to have the final updated weights. 
We used different value loss and policy loss coefficients to balance the weights of two losses. For the Switch games, we used small value loss coefficients because the value loss is between $[0-10]$ while the absolute value policy loss is smaller than $1e-2$. For the Checker game, the value loss and policy loss are at the same scale between $[1e-4, 1e-2]$. Also, we added entropy loss and KL loss to encourage exploration and limit the policy update for each step.  We used~\cite{neng} as the Nash equilibrium solver for finding the meta-game Nash. The Nash solver is CMAES for all the experiments.  
If not particularly indicated, all the baselines use common settings as listed in Table~\ref{tb:hyper}.
  VDN, QMIX use common individual action-value networks as those used by MATRL; each consists of two 128-width hidden layers. 
We includes more experiment result on 4-agent ant task multi-agent MuJoCo task in Fig.~\ref{fig:ant}, which also demonstrate the superior performance of MATRL compared to other settings.
The specialized parameter settings for each algorithm are provided in Table~\ref{tb:hyper_baselines} and~\ref{tb:hyper_baselines_MuJoCo}:

\textbf{Multi-agent Atari Pong}.
The hyper-parameters setting for MATRL are listed in Table~\ref{tb:hyper_baselines_Pong}. We used the same hyper-parameters for MATRL and IL. We take the raw pixel input from the Atari environment, and we processed it with a convolution network, which has filter sizes [8,4,3], kernel sizes (3,3,3), and stride sizes [4,2,1] and "VALID" as padding. Then we pass the processed embedding to a 2 layer fully connected network to get the policy.

\begin{table}[t!]
\caption{MATRL hyper-parameter settings in grid worlds.}
\label{tb:hyper}
\centering
\begin{sc}
\resizebox{1. \linewidth}{!}{
\begin{tabular}{lcl}
\toprule
\textbf{Common Settings }& \textbf{Value} & \textbf{Description}  \\
\midrule
Policy Learning Rate& $0.002$ & Optimizer learning rate.\\
Batch Size &  $2000 $ & Number of data point for each update. \\
Gamma & $0.99$ &  Long Term Discount factor.\\
Hidden Dimension & $128$ & Size of hidden states. \\
Number of Hidden Layers & $2$ & Number of Hidden layers. \\
Nash Equilibrium Solver Method & CMAES & The method for finding the Nash equilibrium of meta-game \\
Neural Network & MLP & The neural network architecture for policy and critic \\
Policy Update Iterations & $10$ & Number of gradient steps for each batch of update.  \\
Best Response Learning Rate &$0.002$ & The learning rate for best response step \\
Best Response Interactions & $5$ &  Number of gradient steps for best response step \\
KL Coefficient & 0.001 & The KL divergence coefficient in calculating loss \\
Entropy Coefficient & 0.05 & The entropy coefficient in calculating loss \\
Policy Ratio Clip  & 0.1 & The clip value for policy ratio \\
Best Response Importance Ratio Clip & 0.1 & The clip value for best response importance weight \\
\hline
\multicolumn{3}{l}{\textbf{2 Player Switch}}  \\  \hline
Value loss Coefficient & 0.01 & The value loss is larger than policy loss\\
\hline 
\multicolumn{3}{l}{\textbf{2 Player Checker}}  \\  \hline
Value loss Coefficient & 1.0 & The value loss is at same scale as policy loss \\
\hline 
\multicolumn{3}{l}{\textbf{4 Player Switch}}  \\  \hline
Value loss Coefficient & 0.01 & The value loss is larger than policy loss \\
\bottomrule
\label{sec:matrl_settings}
\end{tabular}
}
\end{sc}
\end{table}
\begin{table}[t!]
\caption{Hyper-parameter settings for baseline algorithms in grid worlds.}
\label{tb:hyper_baselines}
\centering
\begin{sc}
\resizebox{1. \linewidth}{!}{
\begin{tabular}{lcl}
\toprule
\textbf{Settings }& \textbf{Value} & \textbf{Description}  \\
\midrule 
\multicolumn{3}{l}{\textbf{VDN}}  \\  \hline
Monotone Network Layer    & 2& Layer number of Monotone Network.\\
Monotone Network Size & 128 & Hidden layer size  of Monotone Network. \\
Target Network Update Interval & 200 & Number of iterations between each target network update\\
Learner & Double-Q Learner & The algorithms for each agent \\
\hline
\multicolumn{3}{l}{\textbf{QMIX}}  \\  \hline
joint action-value network layer    & 2& Layer number of joint action-value network.\\
joint action-value network Size & 128 & Hidden layer size of joint action-value network. \\
Learner & Double-Q Learner & The algorithms for each agent \\
\bottomrule
\end{tabular}
}
\end{sc}
\end{table}


\begin{table}[t!]
\caption{Hyper-parameter settings in multi-agent MuJoCo hopper.}
\label{tb:hyper_baselines_MuJoCo}
\centering
\begin{sc}
\resizebox{1. \linewidth}{!}{
\begin{tabular}{lcl}
\toprule
\textbf{Settings }& \textbf{Value} & \textbf{Description}  \\
\midrule 
\multicolumn{3}{l}{\textbf{MATRL and its variants }}  \\  \hline
Agent algorithm & PPO & The learning algorithm for agent \\
Network    & 2 Layer MLP [128, 128] & The network architecture and size for the PPO agent\\
Learning rate & 0.002 & Learning rate for agents \\
Batch ize & 4000 & Batch size for one update\\
Value loss coefficient & 0.001 & the value loss coefficient in total loss \\
Policy loss coefficient & 100 & the policy loss coefficient in total loss \\ 
Policy Update Iterations & $10$ & Number of gradient steps for each batch of update.  \\
Best Response Learning Rate &$0.002$ & The learning rate for best response step \\
Best Response Interactions & $5$ &  Number of gradient steps for best response step \\
Entropy coefficient & 0.05 & the entropy coefficient in total loss \\ 
KL-divergence coefficient & 0.01 & the KL-divergence coefficient in total loss \\
gamma & 0.99 & discount factor \\ 
\hline
\multicolumn{3}{l}{\textbf{MADDPG}}  \\  \hline
Network    & 2 Layer MLP [300, 300] & The network architecture and size for the PPO agent\\
Learning rate & 0.001 & Learning rate for agents \\
Batch size & 100 & Batch size for one update\\
Update interval & 100 & Update the network every 100 time steps \\
Pre-train timeteps & 10000 & Number of time steps before network update \\
gamma & 0.99 & discount factor \\ 
\midrule 
\multicolumn{3}{l}{\textbf{COMIX}}  \\  \hline
Hyper-Network Layer    & 2& Layer number of Hyper-Network.\\
Hyper-Network Size & 64 & Hidden layer size  of Hyper-Network. \\
Act Noise & 200 & Stddev for Gaussian exploration noise added to policy at training time.\\
Learner & Double-Q Learner & The algorithms for each agent \\
\bottomrule
\end{tabular}
}
\end{sc}
\end{table}

\begin{table}[t!]
\caption{Hyper-parameter settings in multi-agent pong Atari.}
\label{tb:hyper_baselines_Pong}
\centering
\begin{sc}
\resizebox{1. \linewidth}{!}{
\begin{tabular}{lcl}
\toprule
\textbf{Settings }& \textbf{Value} & \textbf{Description}  \\
\midrule 
\multicolumn{3}{l}{\textbf{MATRL and its variants }}  \\  \hline
Agent algorithm & PPO & The learning algorithm for agent \\
Network    & 3 Layer CNN, 2 layer FC  & The network architecture and size for the PPO agent\\
Learning rate & 0.002 & Learning rate for agents  \\
Batch ize & 4000 & Batch size for one update\\
Value loss coefficient & 0.1 & the value loss coefficient in total loss \\
Policy loss coefficient & 10 & the policy loss coefficient in total loss \\ 
Policy Update Iterations & $10$ & Number of gradient steps for each batch of update.  \\
Best Response Learning Rate &$0.002$ & The learning rate for best response step \\
Best Response Interactions & $5$ &  Number of gradient steps for best response step \\
Entropy coefficient & 0.05 & the entropy coefficient in total loss \\ 
KL-divergence coefficient & 0.01 & the KL-divergence coefficient in total loss \\
gamma & 0.99 & discount factor \\ 
\bottomrule
\end{tabular}
}
\end{sc}
\end{table}
\end{document}